%% file: main.tex
\documentclass[fleqn,usenatbib,useAMS]{mnras}

\usepackage{newtxtext,newtxmath}
\usepackage[T1]{fontenc}

\DeclareRobustCommand{\VAN}[3]{#2}
\let\VANthebibliography\thebibliography
\def\thebibliography{\DeclareRobustCommand{\VAN}[3]{##3}\VANthebibliography}

\usepackage{graphicx}	% Including figure files

\usepackage{grffile}

\usepackage{amsmath}	% Advanced maths commands
\usepackage{mathtools}
\usepackage{physics}
\usepackage{xcolor}

\input{Planck.tex}

\newlength{\arrow}
\newcommand*{\myrightarrow}[1]{\xrightarrow{\mathmakebox[\arrow]{#1}}}

\title[Planck PR4 NILC $y$-map]{An improved Compton parameter map of thermal Sunyaev-Zeldovich effect from \textit{Planck} PR4 data}

\author[J.~Chandran, M.~Remazeilles, R.~B.~Barreiro]{
Jyothis Chandran,\thanks{E-mail: \href{mailto:chandran@ifca.unica.es}{chandran@ifca.unican.es}}
Mathieu Remazeilles,\thanks{E-mail: \href{mailto:remazeilles@ifca.unican.es}{remazeilles@ifca.unican.es}} 
and R.~B. Barreiro\thanks{E-mail: \href{mailto:barreiro@ifca.unican.es}{barreiro@ifca.unican.es}}  
\\
Instituto de Física de Cantabria (CSIC-UC), Avda. de los Castros s/n, 39005 Santander, Spain
}

\date{Accepted 2023 October 9. Received 2023 October 4; in original form 2023 May 17}

\pubyear{2023}

\begin{document}
\label{firstpage}
\pagerange{\pageref{firstpage}--\pageref{lastpage}}
\maketitle

% Abstract of the paper
\begin{abstract}
 Taking advantage of the reduced levels of noise and systematics in the data of the latest \textit{Planck} release (PR4, also known as NPIPE), we construct a new all-sky Compton-$y$ parameter map (hereafter, $y$-map) of the thermal Sunyaev-Zeldovich (SZ) effect from the \textit{Planck} PR4 data. A tailored Needlet Internal Linear Combination (NILC) pipeline, first validated on detailed sky simulations, is applied to the nine single-frequency \textit{Planck} PR4 sky maps, ranging from $30$ to $857$\,GHz, to produce the PR4 $y$-map over 98\% of the sky. Using map comparisons, angular power spectra and one-point statistics we show that the PR4 NILC $y$-map is of improved quality compared to that of the previous PR2 release. The new $y$-map shows reduced levels of large-scale striations associated with $1/f$ noise in the scan direction. Regions near the Galactic plane also show lower residual contamination by Galactic thermal dust emission. At small angular scales, the residual contamination by thermal noise and cosmic infrared background (CIB) emission is found to be reduced by around 7\%  and 34\%, respectively, in the PR4 $y$-map. The PR4 NILC $y$-map is made publicly available for astrophysical and cosmological analyses of the thermal SZ effect.
\end{abstract}

% Select between one and six entries from the list of approved keywords.
% Don't make up new ones.

%%%%%%%%%%%%%%%%%%%%%%%%%%%%%%%%%%%%%%%% BODY OF PAPER %%%%%%%%%%%%%%%%%%%%%%%%%%%%%%%%%%%%%%%%%%%%%

\section{Introduction}

The cosmic microwave background (CMB) radiation undergoes spectral and spatial distortions while travelling from the last-scattering surface at the recombination epoch up to our instruments at the present time because of the scattering and the deflection of the CMB photons by the matter intervening along the line of sight (los). Such distortions induce secondary CMB temperature anisotropies \citep*{Aghanim2008}. Analysing them allows us to use the CMB as a \emph{backlight} to probe the baryonic and dark matter distributions in the Universe \citep{Backlight2021}.

The most prominent spectral distortion of the CMB anisotropies arises from the thermal Sunyaev-Zeldovich (SZ) effect \citep{1969Ap&SS...4..301Z,1972CoASP...4..173S}: when CMB photons travel through a hot ionised gas of electrons, predominantly located in the potential wells of massive galaxy clusters, they get \emph{upscattered} to higher energies by the electrons through an inverse Compton scattering process. This causes an overall shift of the CMB blackbody spectrum to higher frequencies as the total number of photons is conserved, thus leading to a characteristic spectral signature of the thermal SZ effect,  with a decrement of CMB intensity at low frequency ($<217$\,GHz) in the direction of galaxy clusters and an increment of CMB intensity at high frequency ($>217$\,GHz).   

The peculiar frequency dependence of the thermal SZ effect has allowed the detection of thousands of galaxy clusters from multi-frequency observations of the microwave sky over the past decade \citep{SPT-SZ-Cat2015,planck2014-a36,ACT-SZ-Cat2021,Melin2021}, but also the mapping of the thermal SZ Compton-$y$ parameter (hereafter, $y$-map) from the entire hot gas all over the sky, which includes diffuse, unbound gas between clusters \citep{planck2014-a28,2019A&A...632A..47A,2020PhRvD.102b3534M,2022MNRAS.509..300T,2022ApJS..258...36B}. With increasing sensitivity and resolution, next-generation CMB experiments are expected to release even larger cluster catalogues \citep{CMBS4-2019,SO2019} and cleaner $y$-maps of the hot gas \citep{PICO2019,LB2022} in the near future.

Being independent of the redshift, the thermal SZ effect serves as an important cosmological probe of the large-scale structure in the Universe \citep*{1999PhR...310...97B, 2002ARA&A..40..643C}. Cluster number counts as a function of the redshift from current SZ catalogues provide cosmological constraints on the amplitude of dark matter fluctuations, $\sigma_8$, the matter density, $\Omega_{\rm m}$, and the dark energy equation-of-state parameter, $w$, which are independent of the constraints from primary CMB anisotropies, exhibiting the first tensions with respect to $\Lambda\rm{CDM}$ model predictions from the high-redshift CMB probe \mbox{\citep{planck2013-p15,planck2014-a30}}. Unlike cluster catalogues which solely rely on the most massive clusters that can be detected individually, Compton $y$-maps probe the full thermal SZ emission over the sky, including the fainter emission from low-mass clusters and the diffuse, unbound gas outside clusters which in fact contribute statistically to the signal.  As such, Compton $y$-maps provide another important and complementary cosmological probe through the angular power spectrum of the Compton-$y$ field \citep{Komatsu1999,Refregier2000,Komatsu2002,planck2014-a28,Bolliet2018,Remazeilles2019,Rotti2021,2022MNRAS.509..300T}, higher-order statistics \citep{Rubino-Martin2003,Bhattacharya2012,Wilson2012,Hill2013,planck2014-a28,Remazeilles2019} and cross-correlations with other tracers of the large-scale structure \citep[e.g.][]{Hill2014}.

However, extracting thermal SZ Compton-$y$ anisotropies out of microwave sky observations is challenging because the signal is faint compared to Galactic and extragalactic foreground emissions at submillimetre wavelengths. In addition, thermal noise and instrumental systematics add further contamination to the data. The most significant foreground to thermal SZ emission at small angular scales arises from cosmic infrared background (CIB) anisotropies due to the cumulated emission of dusty star-forming galaxies. At the current stage where the model of various foregrounds like the CIB is relatively poorly known, the use of blind (i.e. non-parametric) component separation methods is warranted for thermal SZ map reconstruction. Hence, the latest all-sky thermal SZ $y$-maps which have been publicly released by the \textit{Planck} Collaboration \citep{planck2014-a28} have been obtained using tailored versions of the blind Internal Linear Combination (ILC) method for the reconstruction of the thermal SZ effect \citep*{2011MNRAS.410.2481R,2013A&A...558A.118H,2013MNRAS.430..370R}. The two public \textit{Planck} thermal SZ maps were named NILC $y$-map and MILCA $y$-map after the respective names of the two component separation methods that were used. 
Both methods are ILC techniques but employ different frameworks for localization in pixel and spherical harmonic domains. For technical details, we refer the reader to \cite{planck2014-a28} and the references therein.

These latest public all-sky thermal SZ $y$-maps date back from 2015 as a product of the second \textit{Planck} PR2 data release \citep{planck2014-a28}. However, the \textit{Planck} mission had four data releases in total, and the latest so-called PR4 data release in 2020 had significant updates with reduced noise and better control of systematics and calibration thanks to the NPIPE processing pipeline \citep{planck2020-LVII}.

In this paper, we reconstruct an updated and improved all-sky thermal SZ Compton $y$-map over 98\% of the sky from the \textit{Planck} Release 4 (PR4) data using a Needlet Internal Linear Combination \citep[NILC,][]{2009A&A...493..835D} specifically tailored for thermal SZ component separation \citep*{2011MNRAS.410.2481R,2013MNRAS.430..370R}. A similar update has recently been reported for the MILCA $y$-map \citep{2022MNRAS.509..300T}, but the $y$-map is not public to our knowledge. Benefiting from the improved quality of the latest \textit{Planck} PR4 data, our new PR4 NILC $y$-map is made public to the community for astrophysical and cosmological SZ analyses and cross-correlation studies. 

This paper is organised as follows. In Section~\ref{sec:data}, we introduce the \textit{Planck} PR4 data used to construct the new $y$-map, as well as some external data sets used as foreground tracers to characterise residual contamination in the $y$-map. In Section~\ref{sec:method}, we describe our implementation of the NILC component separation method for thermal SZ reconstruction, highlighting the differences of processing with respect to the PR2 analysis. We present our results in Section~\ref{sec:validation_map} with a visual inspection of the PR4 NILC $y$-map and comparison with PR2 $y$-maps, estimation of its angular power spectrum and one-point statistics. In Section~\ref{sec:residuals}, we estimate the levels of residual contamination due to foregrounds and noise in the PR4 NILC $y$-map and compare these with those of the PR2 NILC $y$-map. We present our conclusions in Section~\ref{sec:conclusions}.

%%%%%%%%%%%%%%%%%%%%%%%%%%%%%%%%%%%%%%%%%%%%%%%%%%%%%%%%%%%%%%%%%%%%%%%%%%%%%%%%%%%%%%%%%%%%%%%%%%%%
\section{Data} \label{sec:data}

\subsection{\textit{Planck} PR4 data} \label{sec:NPIPE_data}

The latest PR4 (NPIPE) data release from \textit{Planck},\footnote{\url{https://pla.esac.esa.int}} as described in \citet{planck2020-LVII}, is used in this work for thermal SZ Compton $y$-map reconstruction. The NPIPE processing pipeline was used to streamline the conversion of both LFI (Low-Frequency Instrument) and HFI (High-Frequency Instrument) raw time-ordered data (TOD) into nine calibrated full-sky maps corresponding to the nine frequency channels of \textit{Planck}. 

The main differences of the PR4 data with respect to earlier PR2 data that could benefit the reconstructed thermal SZ $y$-map are:
\begin{enumerate}
    \item Reduced noise levels in the PR4 frequency maps due to adding 8\% more data from the repointing manoeuvre. 
    \item A different fitting of 4\,K lines, a better flagging of pixels, and a smoother glitch removal which also contributes to the decrease of noise and half-ring correlations.
    \item A destriping of data done with \texttt{Madam} \citep*{Madam2005} using extremely short baselines which reduces the stripes due to systematic effects in the scanning direction of the \textit{Planck} satellite in the PR4 sky maps.
    \item Differences in the frequency bandpass responses for the PR4 HFI channels due to the differences in calibration between the PR2 processing pipeline and the PR4 NPIPE processing pipeline.
    \item Calibration of LFI and HFI data performed in a coherent pipeline.
\end{enumerate}

We use the nine single-frequency full-mission maps from PR4, ranging from $30$ to $857$\,GHz, to reconstruct the thermal SZ $y$-map. We also use the two half-ring (HR) data splits from PR4 in nine frequency channels, which correspond to the first and second half of each stable pointing period of \textit{Planck} and thus have practically uncorrelated noise. The two data sets from each half-ring are called HR1 maps and HR2 maps from here on. The principal use of this data split is to characterise the statistics of the noise in the PR4 full-mission $y$-map but also to produce additional HR1 and HR2 $y$-maps with maximally uncorrelated noise for thermal SZ power spectrum estimation.

All resultant $y$-maps are given in the \texttt{HEALPix} pixelation scheme\footnote{\url{https://healpix.sourceforge.io/}} \citep{2005ApJ...622..759G} with a pixel resolution of $N_{\rm side} = 2048$. The input PR4 sky maps are of resolution $N_{\rm side} = 1024$ for the LFI frequency channels ($30$-$70$\, GHz) and $N_{\rm side} = 2048$ for the HFI frequency channels ($100$-$857$\, GHz). For each frequency channel map, there is a smoothing effect due to the finite resolution of the optical beam of the detectors. This is treated using an effective symmetric beam transfer function for each channel. The specific instrumental beam windows from PR4 are used for component separation instead of the approximate Gaussian beam windows used in the PR2 $y$-map analysis \citep{planck2014-a28}. The PR4 beam window at $353$\,GHz, for instance, deviates from the Gaussian approximation by approximately $2\%$ on average across the multipole range $\ell = 1000$-$2048$. This discrepancy thus occurs at small angular scales where the thermal SZ signal from galaxy clusters prevails. Beam modelling errors are similar to calibration errors in an ILC, for which previous studies have demonstrated that even a minor percentage error can degrade the signal reconstruction in the high signal-to-noise regimes \citep*{Dick2010}. Therefore, accurate beam deconvolution at $\ell > 1000$ using PR4 instrumental beams, instead of relying on Gaussian approximations, is preferred for the reconstruction of the small-scale features in the thermal SZ signal from compact galaxy clusters.

\subsection{Masks} \label{sec: masks}

\begin{figure}
    \centering
    \includegraphics[width=\columnwidth]{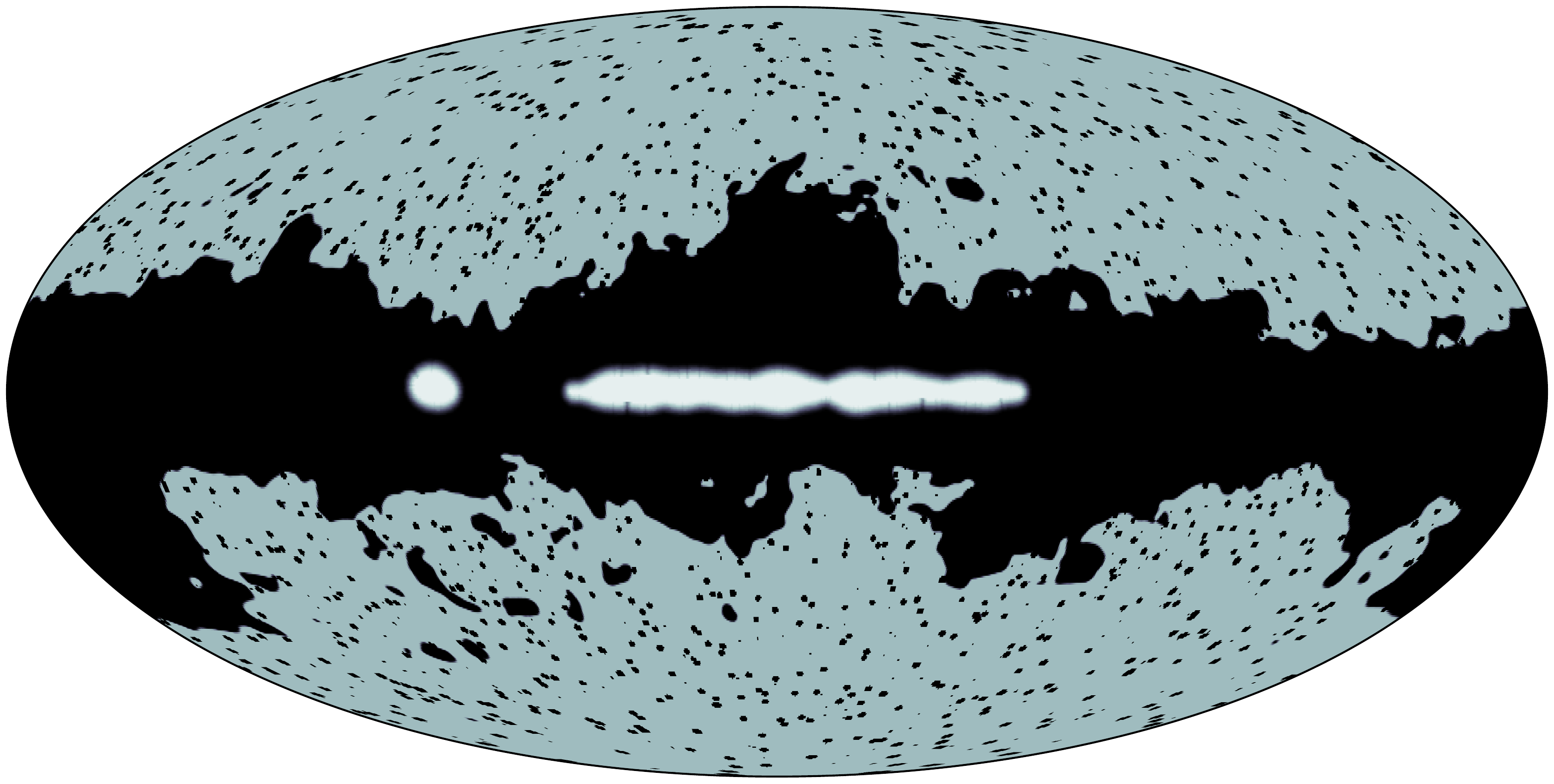}
    \caption{Masks used in the analysis, including the small processing mask for component separation (NILC-MASK, \emph{white area}) leaving $f_{\rm sky}=98\%$ of the sky, the Galactic mask from \citet{planck2014-a28} (GAL-MASK, \emph{black area}) leaving $f_{\rm sky}=60\%$ of the sky and the point-source mask (PS-MASK, \emph{black dots}) for statistical analysis.}
    \label{fig:masks}
\end{figure}

Although the component separation process by NILC is fairly localised on the pixelated sphere by construction, it is not perfectly local, so that the few pixels with the strongest emission in the Galactic centre can create ringing effects when we perform spherical harmonic transforms during needlet decomposition. These ringing effects can result in an overestimation of the sky-RMS signal at higher Galactic latitudes, ultimately affecting the effectiveness of foreground cleaning in those areas. To prevent unwanted ringing effects, we have followed the same strategy used in the \emph{Planck} PR2 analysis \citep{planck2014-a28}, by masking only the brightest 2\% of pixels at $857$\,GHz along the Galactic ridge  in all PR4 frequency maps before passing them through the NILC pipeline. This small processing mask, called NILC-MASK hereafter, is shown in Fig.~\ref{fig:masks} as the white area. The resultant PR4 $y$-map is thus delivered over a fraction $f_{\rm sky}=98\%$ of the sky.

The statistical analysis of the PR4 $y$-map, including estimation of the SZ power spectrum and one-point probability density function (1-PDF) of the $y$-map, requires masking the brightest extragalactic sources and a larger portion of the Galactic region in the $y$-map to mitigate the residual foreground contamination after component separation. We use the apodized Galactic mask released from the PR2 analysis \citep{planck2014-a28}, hereafter called GAL-MASK, conserving about ${f_{\rm sky}=60\%}$ of the $y$-map for statistical analysis. 

For masking extragalactic radio sources in the PR4 $y$-map (see Section~\ref{sec:residual_ps}), we use \textit{Planck} point-source masks specifically constructed for the PR4 data at each frequency channel using the Mexican Hat Wavelet 2 \citep{lopezcaniego2006, planck2014-a35} as part of the Sevem pipeline \citep{planck2020-LVII}. For those frequency channels with higher resolution than $10'$ (i.e. $\ge$ 100~GHz), the masks are further convolved with a Gaussian beam of $10'$  and made binary again by setting a threshold of 0.75. This is because our PR4 $y$-map, like the public PR2 $y$-maps, has a resolution of $10'$, so we need to increase the hole size for the sources to match this resolution. To study the effect of point source residuals in the $y$-map, three different combinations of point-source masks, including only LFI, $30$-$143$\,GHz or all frequency channels are considered in this work (see section \ref{sec:residual_ps}). Our reference point-source mask corresponds to that constructed from the masks of channels $30$ to $143$\,GHz, which we call PS-MASK.
For power spectra computation, the PS-MASK has been apodized with $0.1$ deg transition length using the C1 apodization scheme in \texttt{NaMaster} \citep{2019MNRAS.484.4127A}. 

The combined GAL-MASK and PS-MASK for statistical analysis is displayed in black in Fig.~\ref{fig:masks}, retaining a total sky fraction of 56\%.

\subsection{Foreground tracers} \label{sec:foreground_tracers}

Estimating the residual contamination left by Galactic and extragalactic foregrounds after component separation is an essential part of determining the quality of the thermal SZ $y$-map. This can be done at the map level, by visual comparison of the $y$-map and a foreground template in specific regions of the sky, or by computing the cross-power spectrum between the foreground template and the $y$-map as long as the foreground template does not suffer from thermal SZ contamination. 

The two major foreground contaminants in thermal SZ maps are the CIB, a diffuse extragalactic dust emission from early star-forming galaxies yielding significant power at small angular scales, and the thermal dust emission from our Galaxy which prevails at large angular scales (see Fig.~\ref{fig:tSZ_cl_plk_sims} in Appendix~\ref{app:sims}). 

To assess residual Galactic dust contamination in the $y$-maps, we use as dust template the Improved Reprocessing of the IRAS Survey (IRIS) 100-\micron\ map \citep{1984ApJ...278L...1N, 2005ApJS..157..302M}. At a wavelength of 100 \micron~  (equivalently, a frequency of $\sim 3000$\,GHz), the thermal SZ effect is completely negligible in the IRIS 100-\micron\ map and the sky emission is dominated by Galactic thermal dust emission, making the IRIS 100\micron\ map a reliable tracer of large-scale dust contamination in the $y$-maps.

To assess residual CIB contamination at small angular scales in the $y$-maps,  we use two independent \textit{Planck}-based templates of the CIB emission, both at $857$\,GHz because the \textit{Planck} $857$\,GHz channel map is not used for the construction of both PR2 and PR4 $y$-maps at multipoles $\ell > 300$ \citep[see][and Section~\ref{sec:method}]{planck2014-a28}, which prevents from unwanted noise correlations between the $y$-map and the CIB template. The intensity of the thermal SZ emission is also mostly insignificant at $857$\,GHz in the CIB templates. As a first template, we use the \textit{Planck} GNILC CIB map at $857$\,GHz \citep{planck2016-XLVIII}, which was processed with the data-driven Generalized Neelet ILC (GNILC) method \citep*{2011MNRAS.418..467R} to disentangle CIB from thermal dust emission. As a second, independent template, we use the CIB map at $857$\,GHz from \citet*{Lenz2019}, derived from \textit{Planck} data using a model-dependent approach to subtract thermal dust contamination based on H\textsc{i} gas column density.

%%%%%%%%%%%%%%%%%%%%%%%%%%%%%%%%%%%%%%%%%%%%%%%%%%%%%%%%%%%%%%%%%%%%%%%%%%%%%%%%%%%%%%%%%%%%%%%%%%%%

\section{Methodology} \label{sec:method}

We apply mostly the same NILC algorithm to \textit{Planck} PR4 data as the one used for the PR2 $y$-map release in \citet{planck2014-a28}, with some nuances as the data sets have different characteristics. The major steps and specifications of the current implementation of NILC on the \textit{Planck} PR4 data are described hereafter.

\subsection{Signal modeling} \label{sec:signal_model}

The thermal SZ (tSZ) effect is a frequency-dependent anisotropic distortion of the CMB temperature resulting from inverse Compton scattering of CMB photons off a hot gas of free electrons \citep{1969Ap&SS...4..301Z}:
\begin{align} \label{eq:tsz}
    \Delta T_{\rm tSZ}(\nu, \hat{\mathbf{n}}) = g(\nu)\,y(\hat{\mathbf{n}})\,.
\end{align}

\begin{figure}
    \includegraphics[width=\columnwidth]{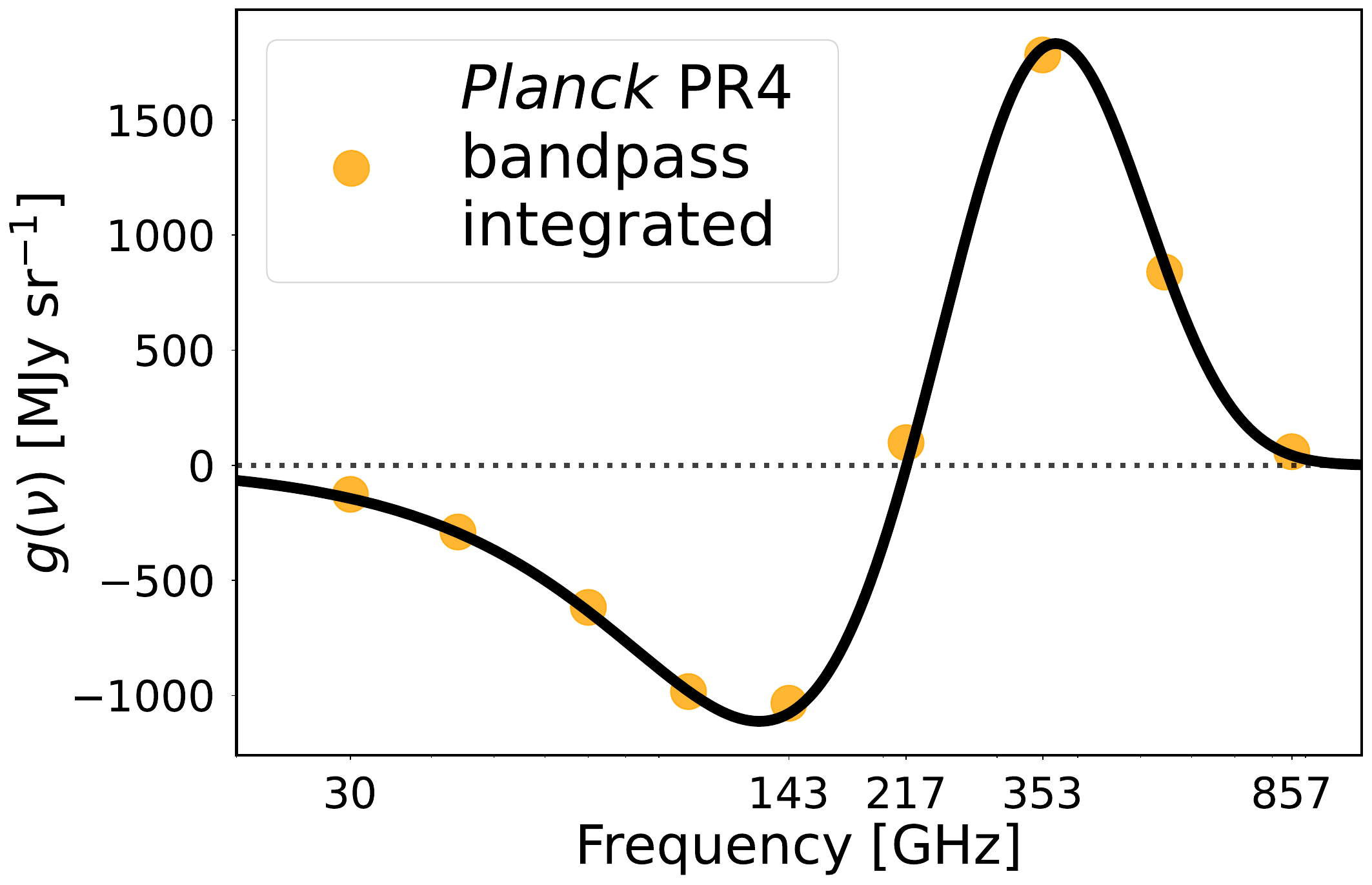}
    \caption{Frequency dependence (SED) of thermal SZ effect in intensity units (\emph{black line}), integrated over \textit{Planck} PR4 channel bandpasses (\emph{yellow dots}).}
    \label{fig:freq_response}
\end{figure}

Here, $g\left(\nu\right)$ is the characteristic frequency-dependence of the distortion  (Fig.~\ref{fig:freq_response}). In the non-relativistic limit and in thermodynamic temperature units, it is given by the analytic form:     
\begin{align} \label{eq:sed} 
 g\left(\nu\right) = T_{\rm CMB}\left( x \coth\left( \frac{x}{2} \right) - 4  \right)\,,
\end{align}
where $x$ is the dimensionless frequency defined as
\begin{align} \label{eq:freq} 
x = \frac{h\nu}{k T_{\rm CMB}}\,,
\end{align}
Here, $h$ is the Planck constant, $k$ is the Boltzmann constant, and $T_{\rm CMB}$ is the CMB blackbody temperature.

The direction-dependent Compton parameter $y(\hat{\mathbf{n}})$ in equation~\eqref{eq:tsz} represents the amplitude of the distortion, which is proportional to the los integral of the electron gas pressure ${P_{\rm e}=n_{\rm e}k T_{\rm e}}$ through the Thomson scattering cross-section:
\begin{align} \label{eq:y}
y\left(\hat{\mathbf{n}}\right) = \frac{ \sigma_T}{m_{\rm e} c^2}\int_{\rm los} n_{\rm e}k T_{\rm e}\left(\hat{\mathbf{n}},l\right) dl\,.
\end{align}
In the equation above, $ T_{\rm e}$ is the electron gas temperature, $n_{\rm e}$ is the electron number density, $m_{\rm e} c^2$ is the electron rest mass energy, and $\sigma_T$ is the Thomson scattering cross-section. 

To get an accurate spectral response of the thermal SZ effect in \textit{Planck} frequency bands, the frequency dependence $g(\nu)$, hereafter spectral energy distribution (SED), must be integrated over \textit{Planck} PR4 frequency bandpasses. The resulting thermal SZ SED coefficients across frequencies for PR4 are shown in Fig.~\ref{fig:freq_response} (yellow dots) and listed in Table~\ref{tab:mixing_vector} in thermodynamic temperature units. They slightly differ from the coefficients of the PR2 analysis in HFI channels \citep{planck2014-a28} due to slightly different HFI bandpasses from the PR4 data release.  

Mirroring PR2 assumptions in \citet{planck2014-a28} for the sake of comparison, we have neglected relativistic corrections to the thermal SZ SED \citep*{Challinor1998,Itoh1998} in the PR4 analysis. However, it has been shown by \citet{Remazeilles2019} that relativistic SZ corrections statistically have an impact on the amplitude of the measured \textit{Planck} $y$-map power spectrum and the skewness of the 1-PDF. Future work will consider the release of another $y$-map from PR4 that accounts for the relativistic SZ effect, which arises due to the variable temperature of the electron gas across the sky.

 \begin{table}
    \centering
     \caption{Thermal SZ SED coefficients in both thermodynamic temperature units and intensity units across \textit{Planck} LFI and HFI channels after \textit{Planck} PR4 bandpass integration.}
     \label{tab:mixing_vector}
     \begin{tabular}{lcc}
        \hline
        Frequency & \multicolumn{2}{c}{g($\nu$)} \\[0.1cm]
        $[\text{GHz}]$ & $[\text{K}_\text{CMB}]$ & $[\text{MJy sr}^\text{-1}]$ \\
        \hline
        30 & -5.3364 & -126.1272 \\
        44 & -5.1782 & -289.5735 \\
        70 & -4.7662 & -616.9667 \\
        100 & -4.0292 & -982.6638 \\
        143 & -2.7801 & -1033.3067 \\
        217 & 0.2037 & 98.4488 \\
        353 & 6.1959 & 1782.2068 \\
        545 & 14.4504 & 839.4226 \\
        857 & 26.3576 & 59.1233 \\
        \hline
     \end{tabular}
    \end{table}

The  observed data $x(\nu, p)$  across the frequencies $\nu$ can thus be expressed for all pixels $p$ like:
\begin{equation}\label{eq:model}
    x(\nu, p) = g(\nu)y(p) + n(\nu,p)\,, 
\end{equation}
where $n(\nu,p)$ is the \emph{unparametrized} nuisance term accounting for all possible foreground emissions and the instrumental noise. The approach is therefore \emph{blind} to the foregrounds because we do not assume a specific spectral model for the foregrounds, whose exact spectral properties are much less known than for the thermal SZ signal. Equation~\eqref{eq:model} can be inverted using the blind NILC method described hereafter to recover the Compton-$y$ parameter in each pixel from the multi-frequency data, with minimum-variance residual contamination from foregrounds and noise.

\subsection{NILC implementation} \label{sec:nilc_method}

There is a vast literature on the technical details of the NILC method \citep{2009A&A...493..835D,2011MNRAS.418..467R,2012MNRAS.419.1163B,2013MNRAS.430..370R,Remazeilles2021,2023MNRAS.525.3117C}. Hence, here we outline the main ingredients of the method and the specificities of our implementation for PR4, highlighting where it differs from the PR2 NILC implementation. 

The \textit{Planck} PR4 frequency maps are provided with the dipole and the frequency-dependent dipole-induced quadrupole included \citep{planck2020-LVII}. Therefore, we first subtracted the dipole and the frequency-dependent quadrupole from the PR4 frequency maps using the best-fit templates from Commander \citep{2008ApJ...676...10E} that are available at NERSC.\footnote{NERSC:/global/cfs/cdirs/cmb/data/planck2020/npipe/commander\_dipole\_\\templates/planck/}
The PR4 maps are also masked with the small NILC-MASK (Fig.~\ref{fig:masks}) to discard the 2\% most-contaminated region in the Galactic plane from the NILC analysis, so that the released PR4 $y$-map effectively covers $f_{\rm sky}=98\%$ of the sky.

In order to produce the PR4 $y$-map at the same $10'$ angular resolution as the public PR2 $y$-maps, the PR4 frequency maps have to be \emph{deconvolved} from their native instrumental beam and \emph{reconvolved} with a common $10'$ symmetric Gaussian beam. In contrast to the PR2 SZ analysis where approximate Gaussian beams were used for beam deconvolution \citep{planck2014-a28}, here we use the specific PR4 beam transfer functions\footnote{The PR4 beam transfer functions are available at NERSC: /global/cfs/cdirs/cmb/data/planck2020/npipe/npipe6v20/quickpol\\/Bl\_npipe6v20\_$\nu$GHzx$\nu$GHz.fits} of each frequency channel to ensure an accurate reconstruction of the small-scale thermal SZ emission from compact clusters. This operation is done in harmonic space and summarised by the following scheme:
\begin{align} \label{eq:beaming}
 x(\nu,p) \myrightarrow{\text{SHT}} x^\nu_{\ell m} \myrightarrow{\times} a^\nu_{\ell m} =  x^\nu_{\ell m} \times \frac{b_\ell^{\,\rm gauss}}{b_\ell^{\,\rm PR4}[\nu]}\,,
 \end{align}
where SHT stands for spherical harmonic transform, $b_\ell^{\,\rm PR4}[\nu]$ are the PR4 beam transfer functions of each channel $\nu$ and $b_\ell^{\,\rm gauss}$ is the $10'$ Gaussian beam transfer function.

Instead of operating on the maps $x(\nu,p)$ in pixel space (pixel-based ILC) or on the spherical harmonic coefficients $a_{\ell m}^\nu$ in harmonic-space (harmonic ILC), NILC operates on \emph{needlet} coefficients. Needlets are a set of wavelets forming a tight frame on the sphere which provide simultaneous localization in pixel domain and in harmonic space \citep*{doi:10.1137/040614359, 2008MNRAS.383..539M}. Localization in the pixel domain is important for the reconstruction of spatially localized signals such as thermal SZ and for adjusting foreground cleaning depending on Galactic latitudes, as Galactic foreground contamination is mostly localized at low latitudes while CIB and noise dominate at high latitudes. Simultaneous localization in harmonic space also enables differentiating between Galactic foregrounds dominating at large angular scales and CIB and noise dominating at small angular scales for customised foreground cleaning. In Appendix~\ref{app:hilc}, we have implemented a harmonic ILC (HILC) on PR4 data to highlight how NILC clearly outperforms HILC on thermal SZ map reconstruction.

\begin{figure}
    	\includegraphics[width=\columnwidth]{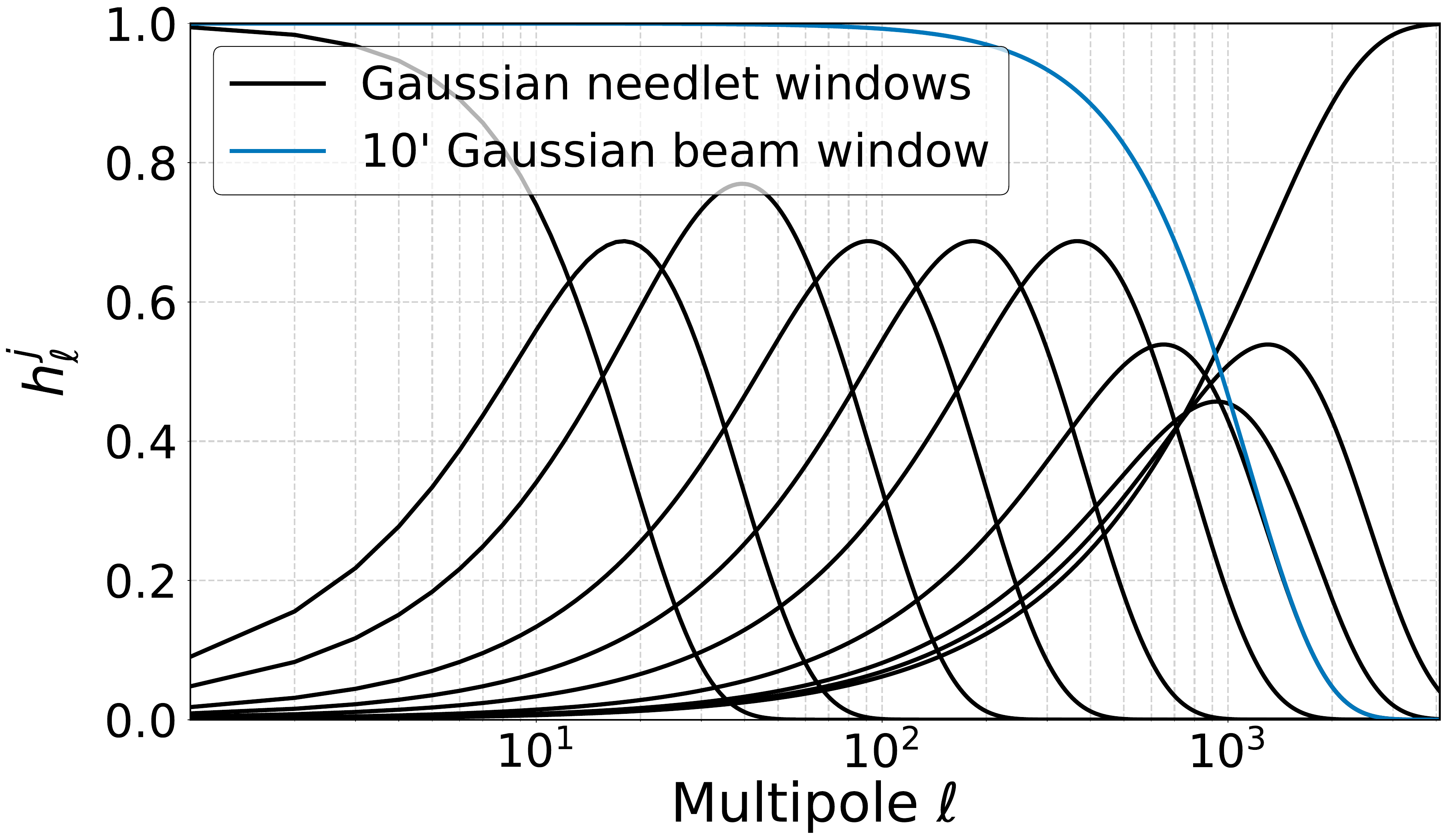}
        \caption{Gaussian-shaped needlet windows (\emph{black}) for localization in angular scales. The output $10'$ beam window of the $y$-map is also shown (\emph{blue}).}
        \label{fig:gauss_bands}
    \end{figure}

To perform the decomposition of the PR4 data into needlet coefficients, we define 10 Gaussian-shaped needlet window functions $h_\ell^j$ in harmonic space as
    \begin{equation}\label{eq:needlets}
        h_\ell^j=
            \begin{cases}
              b_\ell^{\,\rm gauss}(\theta[j]) & \text{if}\ j=1\,, \\[0.2cm]
              \sqrt{\left[b_\ell^{\,\rm gauss}(\theta[j])\right]^2 - \left[b_\ell^{\,\rm gauss}(\theta[j-1])\right]^2} & \text{if}\ j\in [2,9]\,, \\[0.2cm]
              \sqrt{1 - [b_\ell^{\,\rm gauss}(\theta[j-1])]^2} & \text{if}\ j=10\,,
            \end{cases}
    \end{equation}
    where 
 \begin{equation}\label{eq:gauss}
 b_\ell^{\,\rm gauss}(\theta[j]) = e^{-\frac{\ell\left(\ell+1\right)}{2}\left(\frac{\theta[j]}{\sqrt(8\ln2)}\right)^2}
 \end{equation}
  are the SHT of Gaussian functions with \emph{full-width-at-half-maximum} (FWHM) values ($\theta$)~$\in [600, 300, 120, 60, 30, 15, 10, 7.5, 5]$ in arcmin for $j=1,\ldots,9$. The needlet window functions defined in equation~\eqref{eq:needlets} thus satisfy the condition
  \begin{equation} \label{eq:needlet_bands}
    \sum_{j} \left(h_\ell^j\right)^2 = 1\,,
\end{equation}
which guarantees the conservation of the signal at all angular scales after forward and inverse needlet transformations. As shown in Fig.~\ref{fig:gauss_bands}, the 10 needlet windows $h_\ell^j$ operate as bandpasses in harmonic space, each selecting a range of angular scales to ensure localization in harmonic space for component separation. 

%%%%%%%%%%%%%%%%%%%%%%%%%%%%%%%%%%%%%%%%%%
%%%%%%%%%%%%%%%%%%%%%%%%%%%%%%%%%%%%%%%%%%

For each needlet scale $j$, the needlet coefficients $\gamma^{j,\nu}(p)$ are computed from the spherical harmonic coefficients  $a_{\ell m}^\nu$ of the PR4 maps (equation~\ref{eq:beaming}) bandpass-filtered with the needlet window $h_\ell^j$:
\begin{align} \label{eq:needlet_coeff}
        \gamma^{j,\nu}(p) = \sum_{\ell,m}  h_\ell^j\, a_{\ell m}^{\nu}\, Y_{\ell m}(p)\,,
\end{align}
where $Y_{\ell m}(p)$ are spherical harmonics. For each frequency channel $\nu$, we thus obtain 10 needlet maps $ \gamma^{j,\nu}(p)$, each of them displaying sky emission for a specific range of angular scales as selected by the needlet windows.

An estimate of the Compton-$y$ parameter map at needlet scale $j$ is obtained from a weighted linear combination of the PR4 needlet maps across the frequency channels,
\begin{equation} \label{eq:reconstructed_needlet}
    \widehat{y}_j(p)~=~\sum_{\nu} w_j^{\nu}\, \gamma_j^{\nu}(p)\,,
\end{equation}
using the NILC weights:
\begin{equation} \label{eq:needlet_weights}
    w_j^{\nu}(p) = \frac{\sum_{\nu'}~g_{\nu'} \left[C_j^{-1}(p)\right]^{\nu\nu'}}{\sum_{\nu,\nu'}~g_{\nu'} \left[C_j^{-1}(p)\right]^{\nu\nu'}g_{\nu}}\,.
\end{equation}
These weights depend only on $g_{\nu}$, which are the SED coefficients of the thermal SZ effect across the frequency channels (Table~\ref{tab:mixing_vector}), and $C_j^{-1}(p)$, which is the inverse of the empirical covariance matrix of the PR4 data at pixel $p$ and needlet scale $j$, $C_j(p)$, whose elements for all pairs of frequency channels are estimated as
\begin{equation} \label{eq:needlet_cov_matrix}
    C_j^{\nu\nu'}(p) = \sum_{p'} K_j(p,p')\, \gamma_j^{\nu}(p')\, \gamma_j^{\nu'}(p')\,.
\end{equation}
The Gaussian convolution kernel $K_j(p,p')$ defines, for each needlet scale $j$, the effective size of the pixel domain around pixel $p$ over which the product of data $\gamma_j^{\nu}\, \gamma_j^{\nu'}$ for a pair of frequency channels is averaged. For each needlet scale $j$, the FWHM of $K_j(p,p')$ is chosen small enough for localized estimates of the sky covariance in pixel space, but large enough to average as many pixels as possible to minimise empirical chance correlations between signal and contaminants and keep the so-called ILC bias \citep{2009A&A...493..835D} under control.

By construction, the NILC weights (equation~\ref{eq:needlet_weights}) give unit response to the thermal SZ component since $\sum_{\nu} w_j^{\nu}\, g_{\nu} =1$, such that the NILC estimate $\widehat{y}_j(p)$ (equation~\ref{eq:reconstructed_needlet}) recovers the full thermal SZ signal $y$ at needlet scale $j$ without multiplicative error. There is only an additive error to the $y$-estimate due to residual foreground and noise contamination, but this error is kept as small as possible since the NILC weights give, by construction, the minimum-variance solution at the needlet scale $j$.

An important aspect of the analysis is that different subsets of frequency channels are selected depending on the needlet window $j$ for the construction of the NILC weights (equation~\ref{eq:needlet_weights}) and $y$-map estimate (equation~\ref{eq:reconstructed_needlet}) at that needlet scale (see Table~\ref{tab:frequency_needlet_band}). All nine LFI and HFI channels ($30$-$857$\,GHz) are used by NILC in the first three needlet windows for large angular scales. However, due to their low resolution and limited sensitivity at smaller angular scales, LFI channel maps are excluded from subsequent needlet windows, as they are not well-suited for probing sky emission at these smaller angular scales. Therefore, only the 6 HFI channels ($100$-$857$\,GHz) are combined by NILC in the fourth up to the sixth needlet window. Following \citet{planck2014-a28}, the $857$\,GHz channel map is not used at multipoles $\ell > 300$ to mitigate CIB and infrared source contamination, which can be substantial in this channel at small angular scales. Consequently, only 5 HFI channels ($100$-$545$\,GHz) are combined by NILC in the seventh up to the last needlet band. 

\begin{table}
    \centering
     \caption{Frequency channels used for each needlet band.}
     \label{tab:frequency_needlet_band}
    \begin{tabular}{l |c c c c c c c c c r}
    & \multicolumn{10}{c}{Needlet band}  \\
    Frequency  & 1 & 2 & 3 & 4 & 5  & 6 & 7 & 8 & 9 & 10\\
    \hline
    30 GHz   & \checkmark & \checkmark & \checkmark & X & X & X & X & X & X & X  \\
    44 GHz   & \checkmark & \checkmark & \checkmark & X & X & X & X & X & X & X  \\
    70 GHz   & \checkmark & \checkmark & \checkmark & X & X & X & X & X & X & X  \\
    100 GHz  & \checkmark & \checkmark & \checkmark & \checkmark & \checkmark & \checkmark & \checkmark & \checkmark & \checkmark & \checkmark  \\
    143 GHz  & \checkmark & \checkmark & \checkmark & \checkmark & \checkmark & \checkmark & \checkmark & \checkmark & \checkmark & \checkmark  \\
    217 GHz  & \checkmark & \checkmark & \checkmark & \checkmark & \checkmark & \checkmark & \checkmark & \checkmark & \checkmark & \checkmark  \\
    353 GHz  & \checkmark & \checkmark & \checkmark & \checkmark & \checkmark & \checkmark & \checkmark & \checkmark & \checkmark & \checkmark  \\
    545 GHz  & \checkmark & \checkmark & \checkmark & \checkmark & \checkmark & \checkmark & \checkmark & \checkmark & \checkmark & \checkmark  \\
    857 GHz  & \checkmark & \checkmark & \checkmark & \checkmark & \checkmark & \checkmark & X & X & X & X  \\
    \end{tabular}
\end{table}
    
The reconstructed needlet $y$-maps, $\widehat{y}_j(p)$ (equation~\ref{eq:reconstructed_needlet}), from each needlet scale $j$ are finally transformed into spherical harmonic coefficients,  $\widehat{y}_{\ell m,j}$, and combined together to form the complete PR4 NILC $y$-map, $\widehat{y}(p)$:
\begin{align} \label{eq:needlet_to_real_signal}
\widehat{y}_j(p)  \myrightarrow{\text{SHT}}  \widehat{y}_{\ell m,j}  \myrightarrow{} \widehat{y}(p) = \sum_{\ell,m} \left( \sum_j \widehat{y}_{\ell m,j}\,h_\ell^j  \right) Y_{\ell m}(p)\,.
\end{align}

The same NILC weights, computed from the full PR4 data in Eqs.~\eqref{eq:needlet_weights}-\eqref{eq:needlet_cov_matrix}, are also applied to the two sets of half-ring PR4 maps (HR1 and HR2 frequency maps), which went through the same needlet decomposition process. The resulting PR4 HR1 and HR2 $y$-maps, $\widehat{y}^{\,\text{HR1}}(p)$ and $\widehat{y}^{\,\text{HR2}}(p)$, have same thermal SZ signal and residual foreground contamination but mostly-uncorrelated noise. The half-difference between the PR4 HR1 and HR2 $y$-maps, ${\left(\widehat{y}^{\,\text{HR1}}(p) - \widehat{y}^{\,\text{HR2}}(p)\right) / 2}$, cancels out any sky emission but not the instrumental noise, and thus serves as a noise map estimate whose statistical properties are the same as those of the actual residual noise in the full PR4 $y$-map. In addition, because of least-correlation of noise between the two half-rings, the cross-power spectrum between the PR4 HR1 $y$-map and the PR4 HR2 $y$-map enables to estimate the recovered thermal SZ angular power spectrum corrected for the instrumental noise bias.

\section{PR4 thermal SZ \lowercase{$y$}-map characterization} \label{sec:validation_map}

In this section, we assess the quality of the PR4 NILC $y$-map and compare it with that of the PR2 $y$-maps in terms of noise, residual systematics and foreground contamination, through map inspection, one-point statistics and power spectrum analysis.

\subsection{Maps inspection} \label{sec:maps_inspection}

\begin{figure*}
    \centering
    \includegraphics[width=\textwidth]{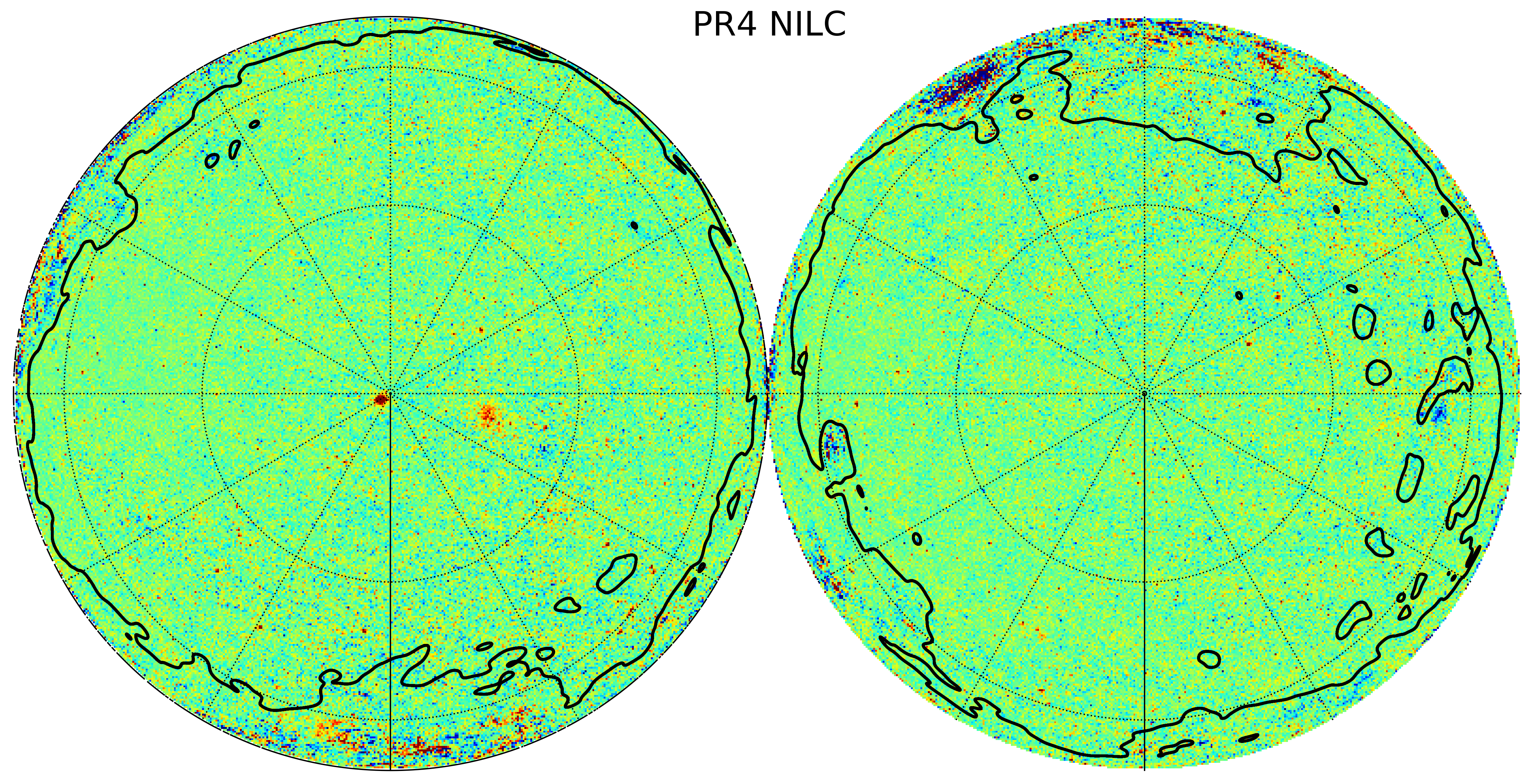}
    \includegraphics[width=\textwidth]{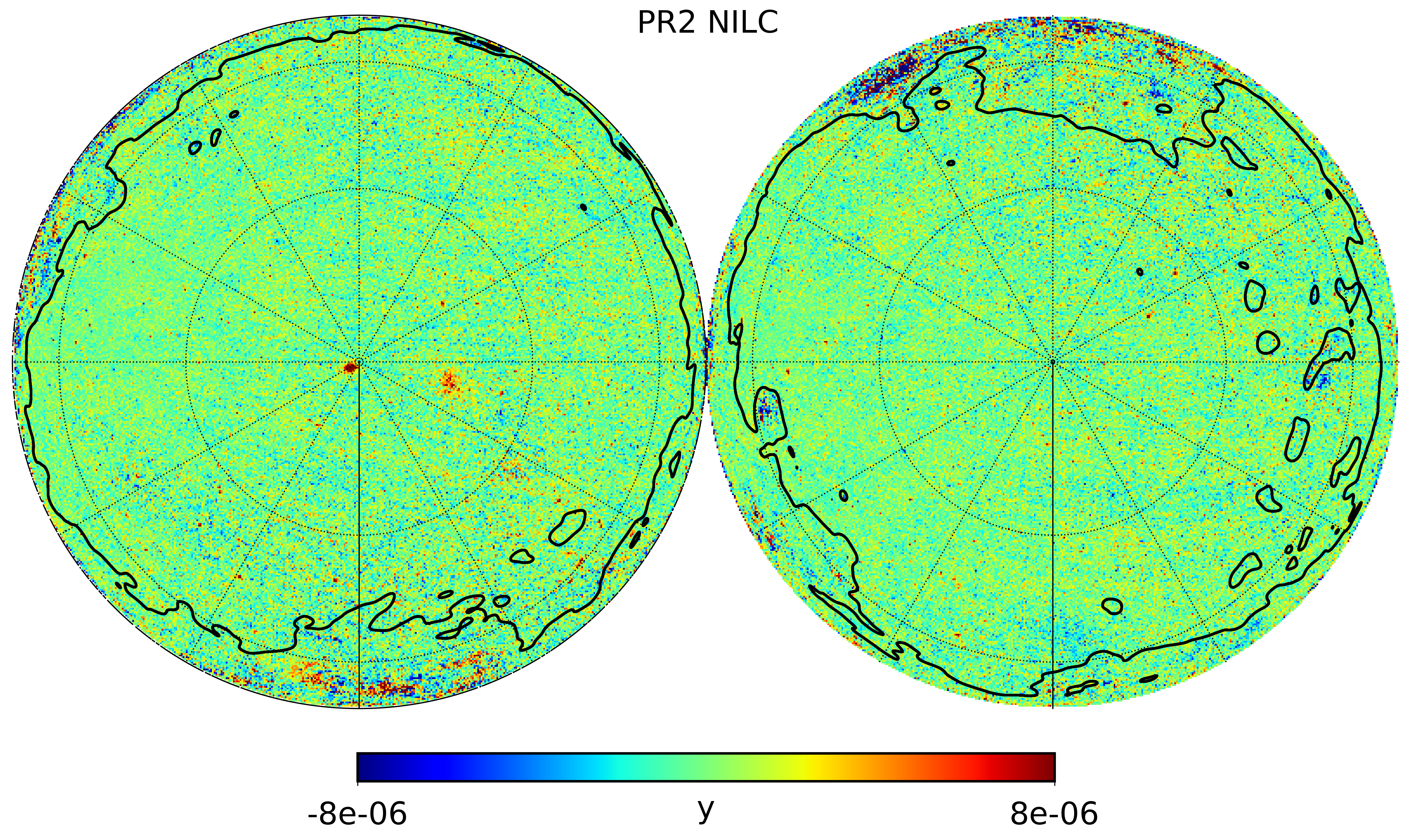}
    \caption{NILC thermal SZ Compton $y$-map from \textit{Planck} PR4 data (\emph{top}; this work) and \textit{Planck} PR2 NILC $y$-map \citep[\emph{bottom;}][]{planck2014-a28}. The left-hand and right-hand sides correspond to the northern and southern hemispheres, respectively, in Galactic coordinates. The black outline shows the boundary of the Galactic mask used for power spectrum analysis.  }
    \label{fig:tSZ_orth_map_plk_npipe_pr2}
\end{figure*}

\begin{figure*}
    \centering
    \includegraphics[width=0.790\textwidth]{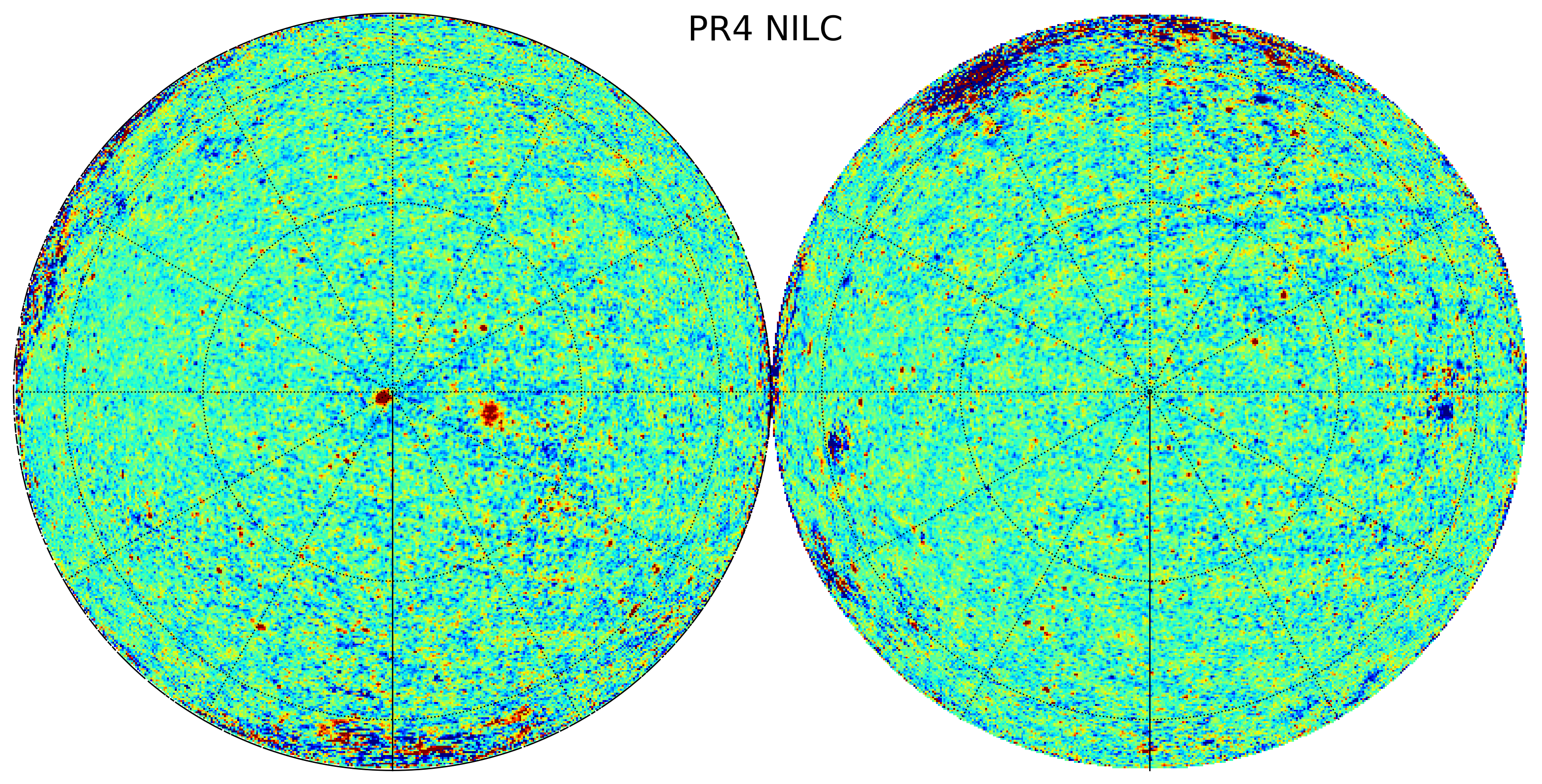}
    \includegraphics[width=0.790\textwidth]{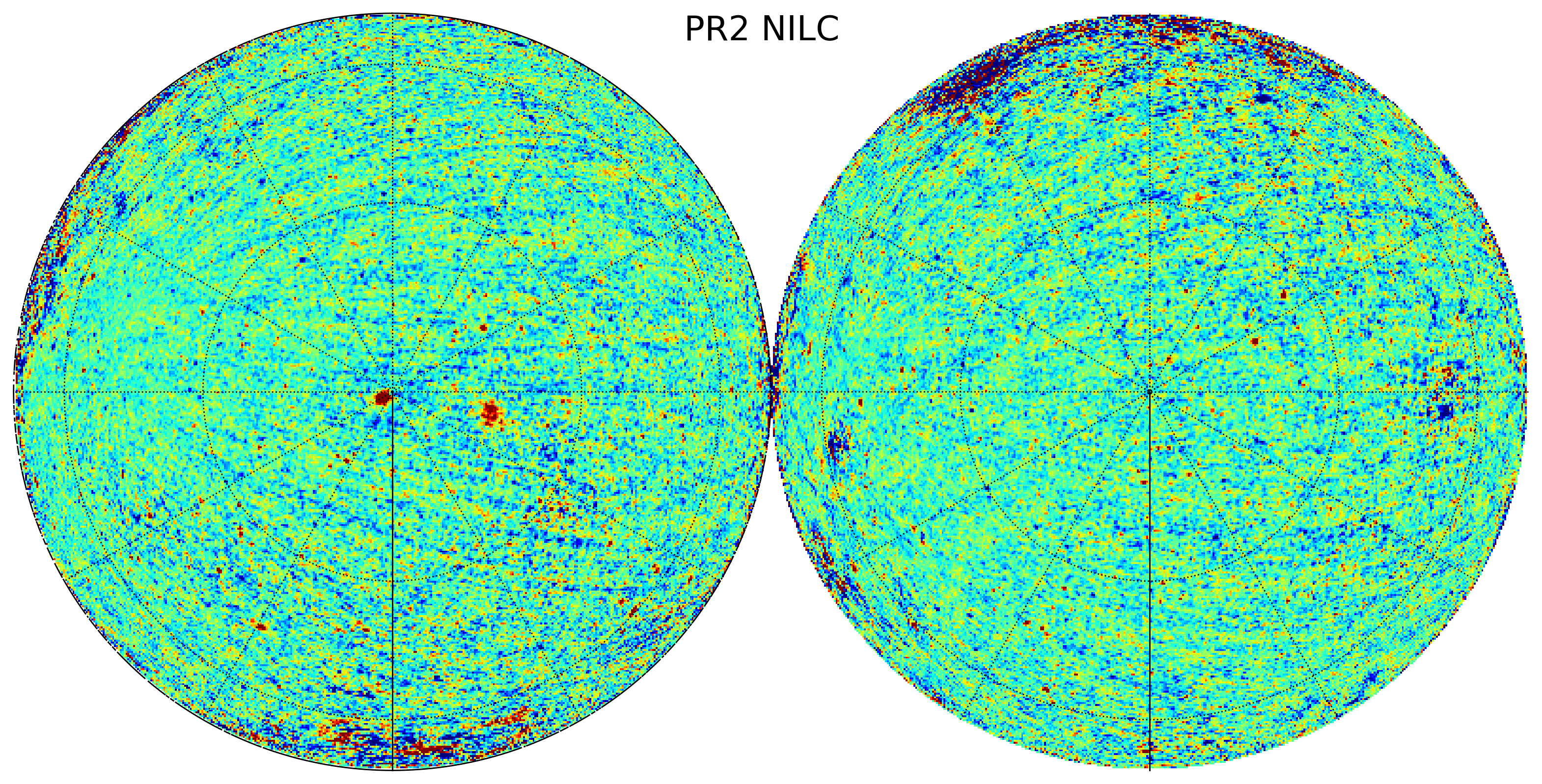}
    \includegraphics[width=0.790\textwidth]{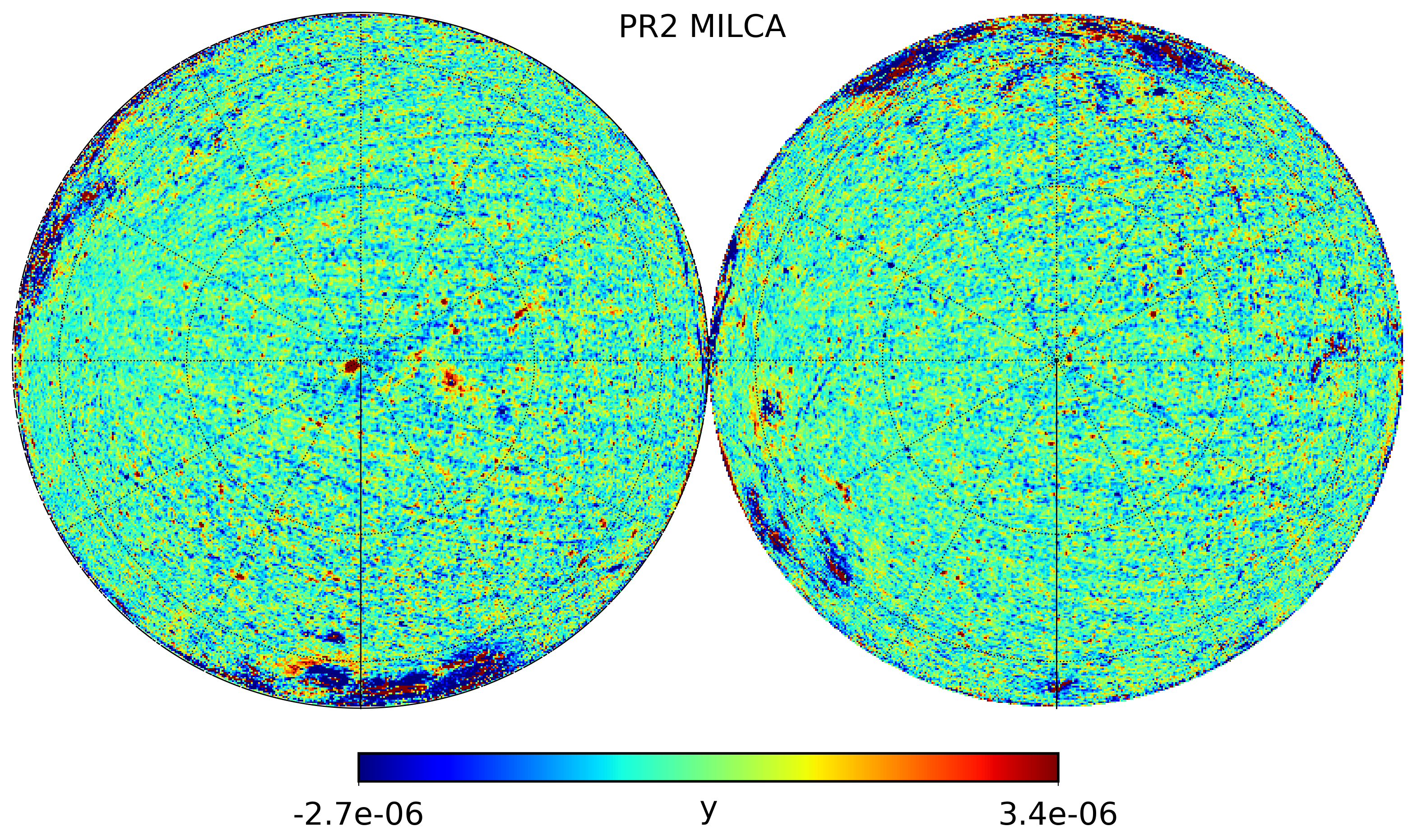}
    \caption{Bandpass-filtered $y$-maps highlighting large-scale stripes along the scan direction due to residual $1/f$ noise: PR4 NILC $y$-map (\emph{top}), PR2 NILC $y$-map (\emph{middle}) and PR2 MILCA $y$-map (\emph{bottom}). The new PR4 NILC $y$-map shows reduced levels of striping compared to the two public PR2 $y$-maps.}
    \label{fig:tSZ_orth_map_plk_npipe_pr2_filtered_asymmetric}
\end{figure*}
 
Fig.~\ref{fig:tSZ_orth_map_plk_npipe_pr2} shows, in orthographic projection, the new PR4 NILC $y$-map (top panel) produced in this work, along with the PR2 NILC $y$-map (bottom panel) released by the \textit{Planck} Collaboration in 2015 \citep{planck2014-a28} for comparison. The left-hand side shows the northern hemisphere and the right-hand side shows the southern hemisphere with respect to Galactic coordinates. Both $y$-maps are at the same $10'$ angular resolution and show relatively high consistency at the map level. However, some differences between the two maps are already visible. The PR2 $y$-map shows more prominent noise (small-scale granular pattern) than the PR4 $y$-map in the bottom part of the northern hemisphere. In addition, blue patchy patterns are visible in the bottom part of the southern hemisphere of the PR2 $y$-map, while these residuals are absent from the new PR4 $y$-map. Thermal SZ sources are visible as red spots in the given colour scale. Prominent galaxy clusters like Coma and Virgo are clearly visible near the north pole. Some residual infrared compact sources may also appear in red in the maps while residual radio sources appear in blue due to the sign of the NILC weights flipping from positive at high frequencies to negative at low frequencies to match the spectral response of the thermal SZ sources. 
Diffuse Galactic foreground contamination, dominated by thermal dust, is visible with significant power at both ends of the range near the Galactic plane (along the edges in Fig.~\ref{fig:tSZ_orth_map_plk_npipe_pr2}).

The black outline in Fig.~\ref{fig:tSZ_orth_map_plk_npipe_pr2} traces the boundary of the GAL-MASK masking $40\%$ of the sky that, together with those pixels contaminated by point sources, are excluded in the 1-PDF and power spectrum analysis (Sections~\ref{sec:pdf} and \ref{sec:C_l}). The regions around the Galactic plane are best to be excluded from statistical analysis due to significant power from residual Galactic emission. However, these regions are still clean enough to spot thermal SZ sources, hence the release of the $y$-map over 98\% of the sky. 

 \begin{figure}
    \centering
    \includegraphics[width=1.0\columnwidth]{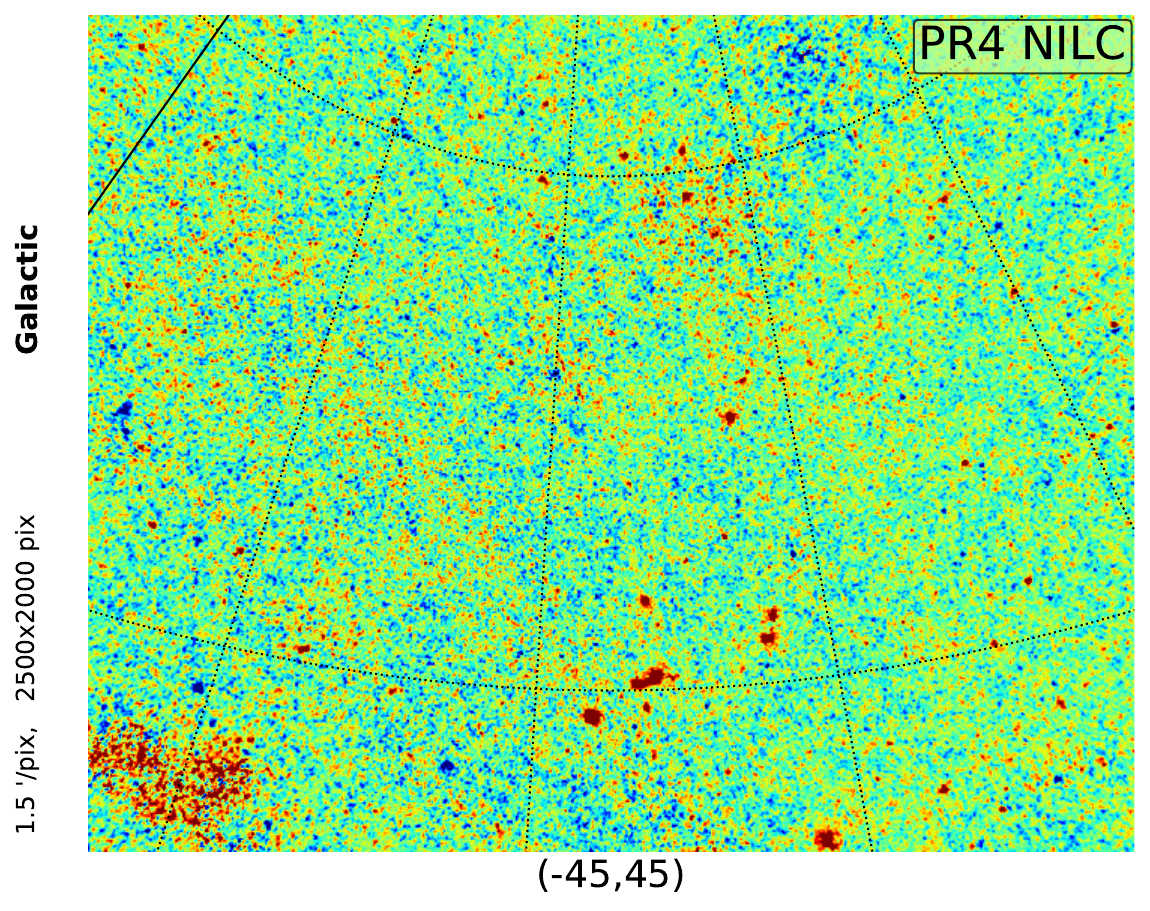}
    \includegraphics[width=1.0\columnwidth]{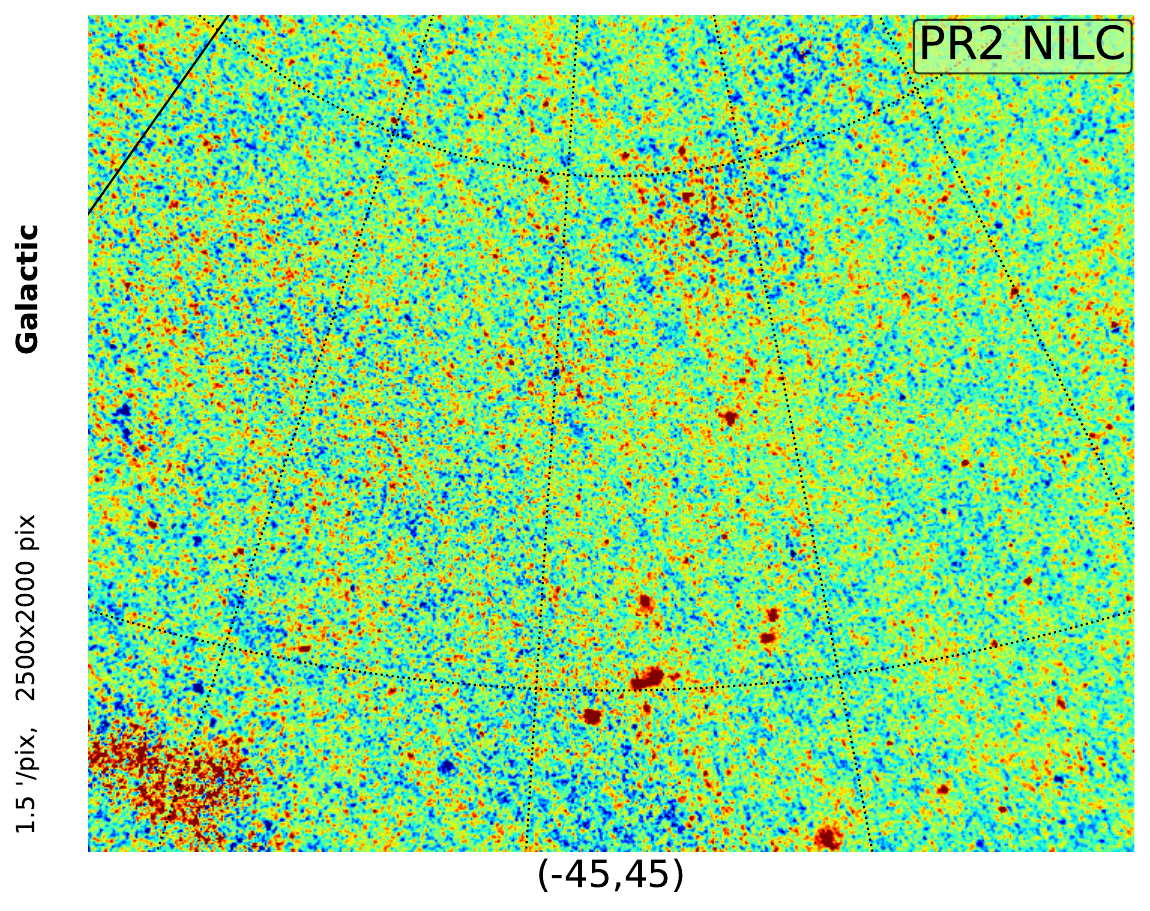}
    \includegraphics[width=1.0\columnwidth]{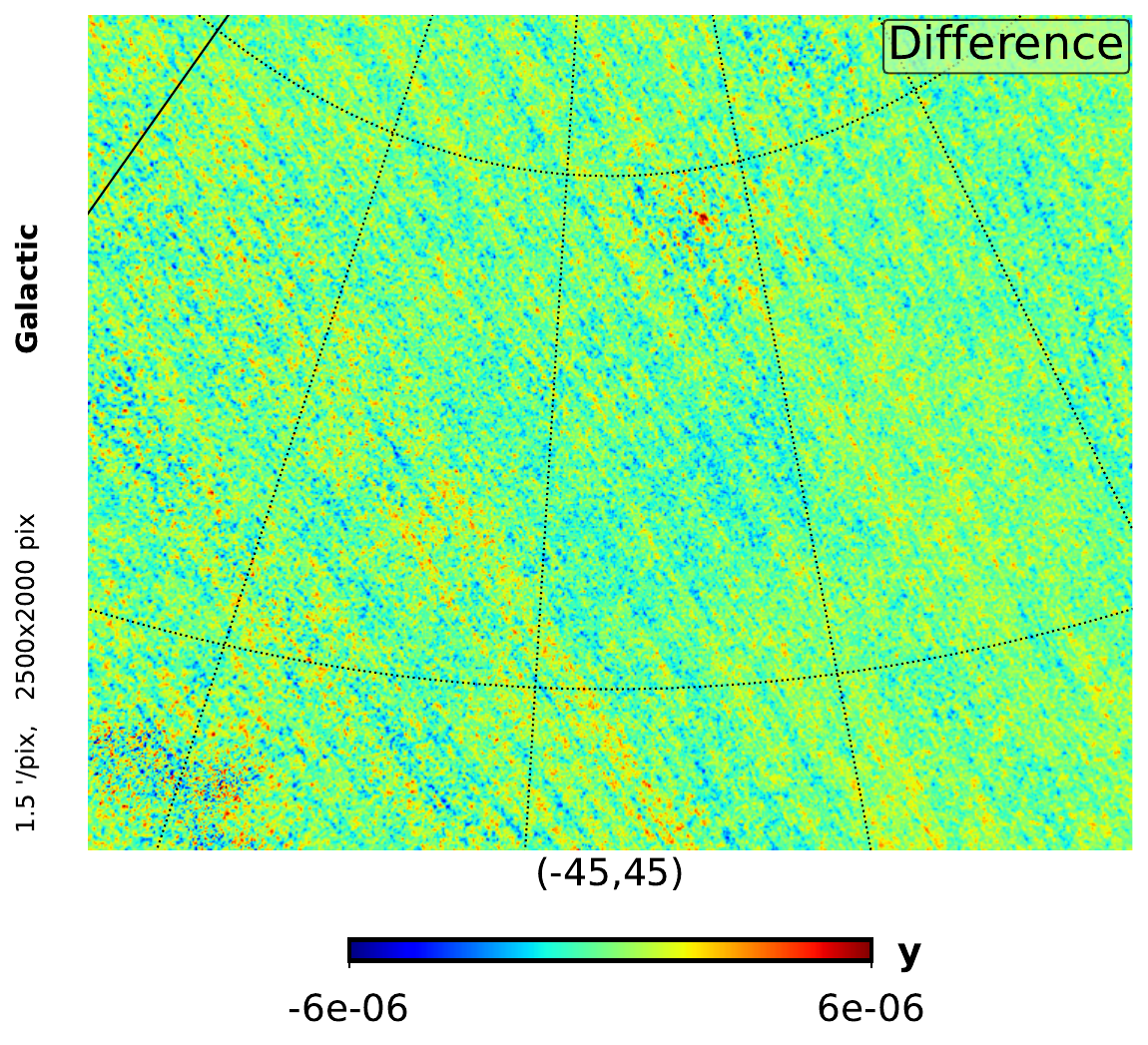}
    \caption{A region at intermediate Galactic latitude in the thermal SZ $y$-maps from \textit{Planck} PR4 (\emph{top}) and PR2 (\emph{middle}) data and their difference $\text{PR4}-\text{PR2}$ (\emph{bottom}). The improved destriping in the PR4 NILC $y$-map compared to the PR2 NILC $y$-map is visible.}
    \label{fig:tSZ_map_plk_npipe_pr2_diff}
 \end{figure}

There are large-scale stripes along the scan direction of the satellite in the \textit{Planck} $y$-maps due to residual systematics in the \textit{Planck} map-making pipeline (see Section \ref{sec:NPIPE_data}). Different methods used for baseline noise correction and destriping affect the morphology of the residual stripes in the $y$-maps \citep[see][section 4.1]{planck2014-a28}. 
Improved destriping in \textit{Planck} PR4 data compared to \textit{Planck} PR2 data is clearly visible in Fig.~\ref{fig:tSZ_orth_map_plk_npipe_pr2_filtered_asymmetric}, which shows the PR4 NILC $y$-map (top) versus the PR2 NILC $y$-map (middle) and the PR2 MILCA $y$-map (bottom) after bandpass-filtering in multipole range. 
All the maps in Fig.~\ref{fig:tSZ_orth_map_plk_npipe_pr2_filtered_asymmetric} have been filtered with a common, analytical bandpass filter which selects the range of multipoles between $\ell=20$ and $\ell=500$, where the thermal SZ signal dominates.
 
Both the PR2 MILCA and PR2 NILC $y$-maps exhibit more residual striping than the PR4 NILC $y$-map. The NILC $y$-maps also show lower levels of residual Galactic foreground contamination around the Galactic plane compared to the PR2 MILCA $y$-map. 
Although the MILCA  $y$-map was updated using PR4 data in \citet{2022MNRAS.509..300T}, it is not publicly available, hence not included in this comparison. While stripes are large-scale residual patterns, they cause the clusters and granular noise in their direction to appear brighter (negative or positive) than the rest, hence the importance of improved destriping in the PR4 $y$-map. 
Fig.~\ref{fig:tSZ_map_plk_npipe_pr2_diff} further highlights the reduced level of destriping at intermediate Galactic latitude in the PR4 NILC $y$-map compared to the PR2 NILC $y$-map.

\begin{figure*}
    \includegraphics[width=\textwidth]{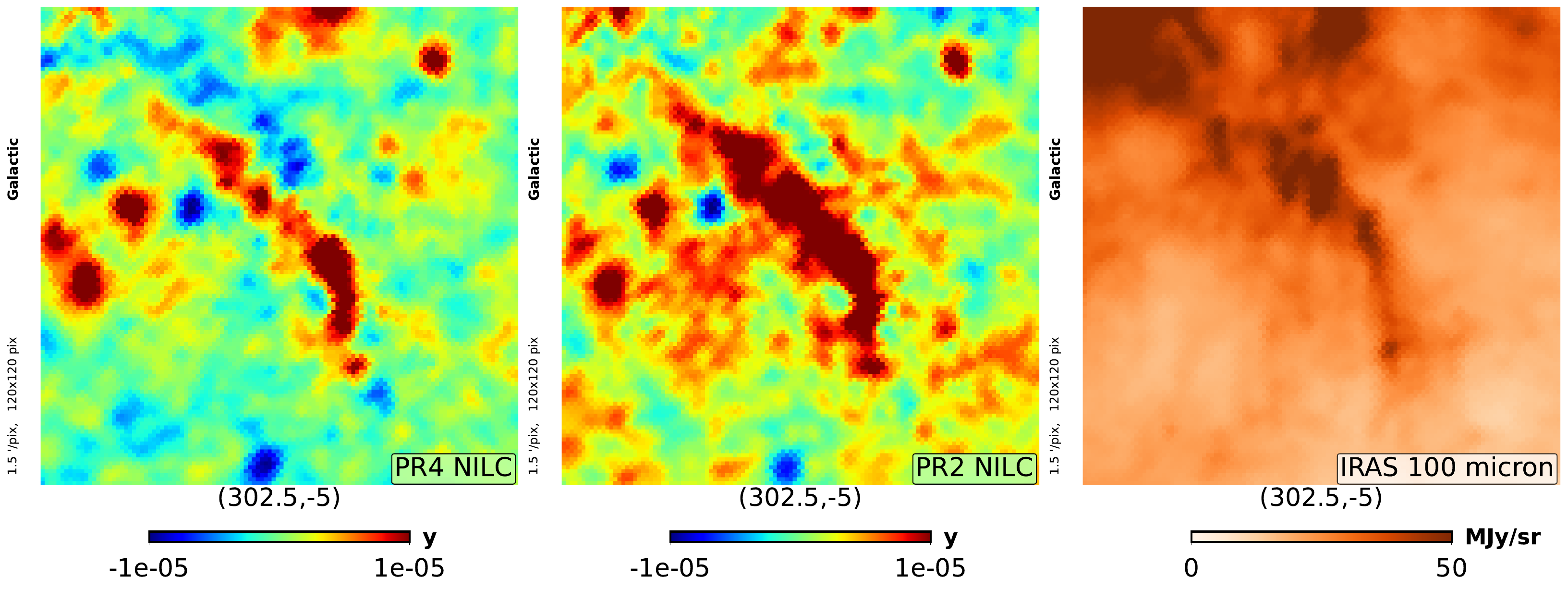}
    \caption{A $3^\circ\times 3^\circ$ region near the Galactic plane, centred at $(\ell,b)=(302.5^\circ,-5^\circ)$, for the PR4 NILC $y$-map (\emph{left}), the PR2 NILC $y$-map (\emph{middle}) and the IRAS 100-micron map (dust tracer, \emph{right}). The PR4 $y$-map is less contaminated by Galactic dust than the PR2 $y$-map.
    }
    \label{fig:tSZ_NPIPE_PR2_dust_noise}
\end{figure*}

Galactic thermal dust emission is the dominant foreground contaminant of the $y$-map at large angular scales and low Galactic latitudes, while extragalactic cosmic infrared background (CIB) emission and instrumental noise are the major contaminants at small angular scales (see e.g. Fig.~\ref{fig:tSZ_cl_plk_sims} from simulations). Local improvements are visible in some regions around the Galactic plane in the PR4 NILC $y$-map where the intensity of the residual Galactic dust emission is reduced compared to the PR2 $y$-map. As an illustration, Fig.~\ref{fig:tSZ_NPIPE_PR2_dust_noise} shows 
one of these regions for both PR4 (left-hand panel) and PR2 (middle panel) $y$-map. The IRIS (reprocessed IRAS) 100-micron map is also shown in the right-hand panel, as a dust tracer, showing some filamentary structure specific to thermal dust emission. It becomes apparent that the contamination is reduced in the PR4 $y$-map with respect to the PR2 version.

\begin{figure}
    \includegraphics[width=\columnwidth]{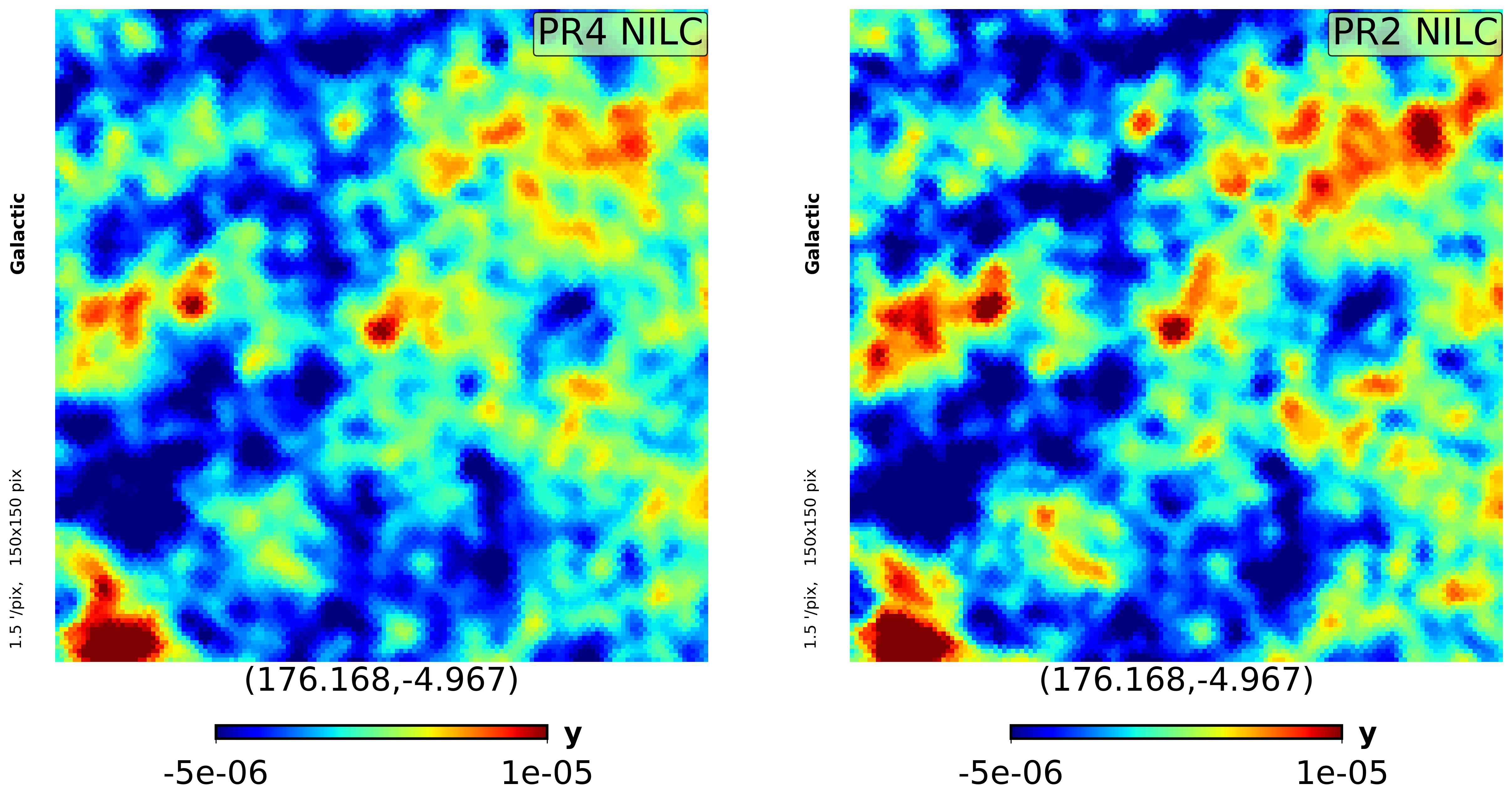}
    \includegraphics[width=\columnwidth]{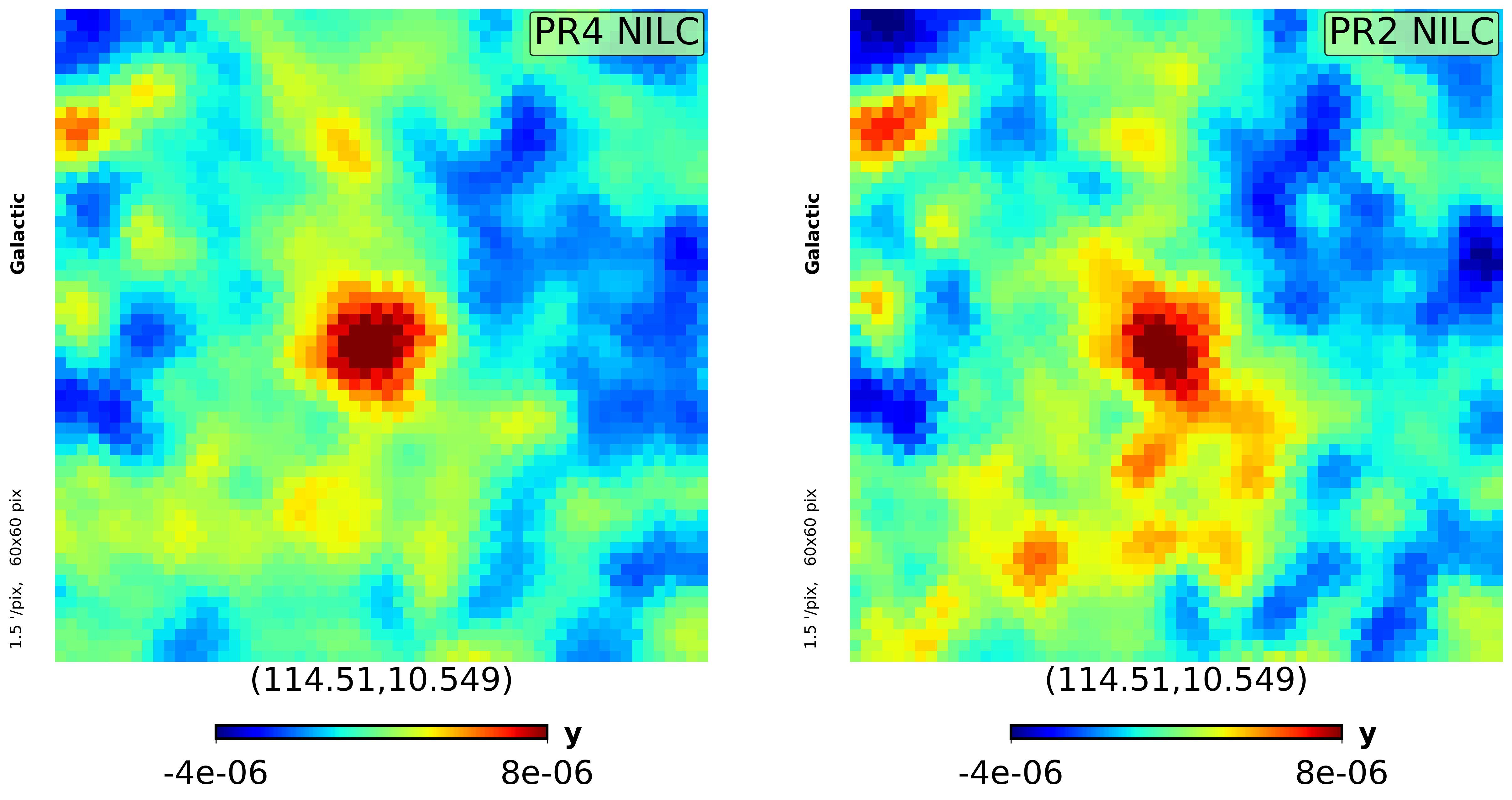}
    \caption{Regions near the Galactic plane centred around galaxy clusters for the PR4 NILC $y$-map (\emph{left}) and the PR2 NILC $y$-map (\emph{right}). The clusters are CIZA J0516.9+2925 (\emph{top}; $4^\circ\times 4^\circ$ patch) and CIZA J2302.7+7137 (\emph{bottom}; $1.5^\circ\times 1.5^\circ$ patch). The diffuse filamentary structure of residual Galactic foreground emission, more prominent in the PR2 $y$-map than in the PR4 $y$-map, alters the morphology of the compact thermal SZ sources.}
    \label{fig:tSZ_NPIPE_PR2_clusters}
\end{figure}

Within the entire Galactic region enclosed by the black line in Fig.~\ref{fig:tSZ_orth_map_plk_npipe_pr2}, the root-mean-square (RMS) of the PR4 $y$-map is reduced by 5\% in comparison to that of the PR2 $y$-map, as a result of diminished Galactic contamination. Extra data and better calibration caused lower noise and systematics in \textit{Planck} PR4 data \citep{planck2020-LVII}, which allows NILC to clean other contaminants like dust (Figs.~\ref{fig:tSZ_NPIPE_PR2_dust_noise}-\ref{fig:tSZ_NPIPE_PR2_clusters}) and CIB (see Section~\ref{sec:residual_cib}) more efficiently in regions and angular scales where they are dominant. This can make thermal SZ sources in the background more visible and identifiable. 

Fig.~\ref{fig:tSZ_NPIPE_PR2_clusters} shows galaxy clusters in regions around the Galactic plane, for which the PR4 $y$-map (left-hand panels) shows lower diffuse foreground contamination than the PR2 $y$-map (right-hand panels). The identified clusters shown in the centre of these images have been observed by the Clusters in the Zone of Avoidance (CIZA) project, which is an X-ray survey for galaxy clusters hidden by the Milky Way \citep{2002ApJ...580..774E, 2007ApJ...662..224K}. 
Their CIZA names are given in the caption of Fig.~\ref{fig:tSZ_NPIPE_PR2_clusters}. These regions should be excluded for cosmological inference as the contamination from Galactic foregrounds is still strong enough to cause biases on the thermal SZ power spectrum. However, lower residuals from the diffuse and filamentary structure of the Galactic foreground emission in the PR4 $y$-map could in principle help in identifying more thermal SZ sources in regions near the Galactic plane and in defining their morphology with better accuracy.

\subsection{1-PDF analysis} \label{sec:pdf}

When CMB photons pass through hot gas in galaxy clusters, they are more likely to gain energy by inverse Compton scattering than to lose it. Therefore, the Compton-$y$ parameter field has a highly non-Gaussian distribution with a positive skewness \citep{Rubino-Martin2003}. Being mostly insensitive to Gaussian contaminants, the characteristic skewness of the thermal SZ emission can also be used, as an advantageous alternative to the power spectrum, to constrain the cosmological parameter $\sigma_8$ \citep{Rubino-Martin2003,Bhattacharya2012,Wilson2012,Hill2013,planck2014-a28,Remazeilles2019}. 

\begin{figure}
	\includegraphics[width=\columnwidth]{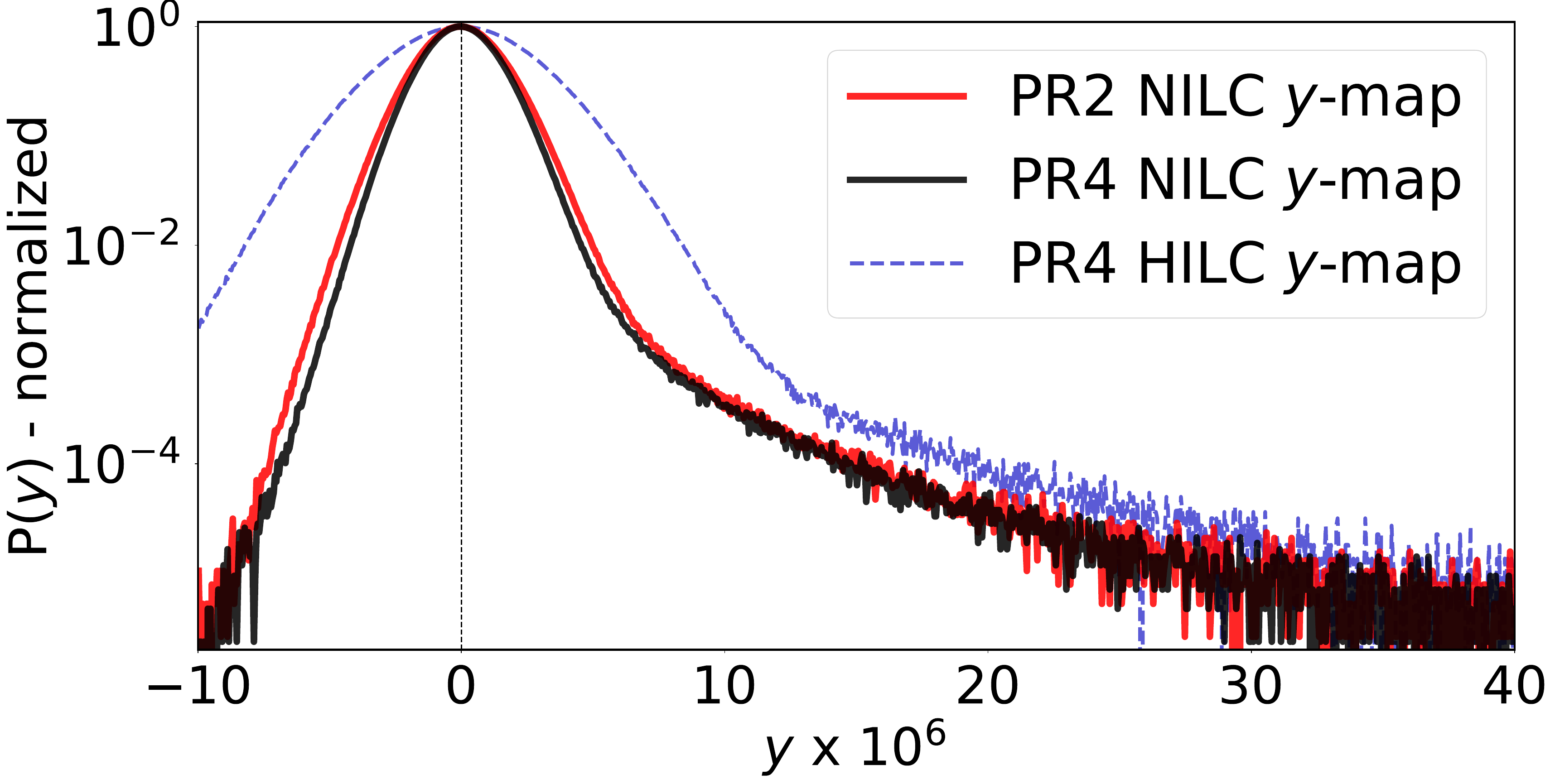}
    \caption{Normalised 1-PDF of  \textit{Planck} thermal SZ Compton $y$-maps: PR4 NILC $y$-map (\emph{solid black}; this work), PR2 NILC $y$-map \citep[\emph{solid red};][]{planck2014-a28} and PR4 HILC $y$-map (\emph{dashed blue}). 
    }
    \label{fig:tSZ_1PDF}
\end{figure}

The one-point \textit{probability distribution function} (1-PDF) of the thermal SZ Compton-$y$ field is computed from the histogram of the $y$-maps over 56\% of the sky after masking the Galactic plane with the GAL-MASK and the point-sources with the PS-MASK. The resulting 1-PDF of the PR2 (solid red) and PR4 (solid black) $y$-maps, normalised to unity at maximum, are plotted in Fig.~\ref{fig:tSZ_1PDF}, showing the positive, skewed tail of the distribution which is characteristic of thermal SZ emission. 
The positive tails of the PR2 and PR4 NILC $y$-maps match together, ensuring consistent recovery of the thermal SZ signal. However, the widths of the distributions do not match, indicating a reduced variance by $17\%$ due to the lower foreground and noise contamination in the PR4 NILC $y$-map compared to the PR2 NILC $y$-map. Reduced variance owing to lower noise in the PR4 NILC $y$-map is also evident from the tighter 1-PDF of the noise (see Fig.~\ref{fig:tSZ_noise_1PDF} and Section~\ref{sec:residual_noise}), which was computed for PR2 and PR4 from the histogram of the half-difference of HR1 and HR2 $y$-maps. 

The blue dashed line in Fig.~\ref{fig:tSZ_1PDF} shows the 1-PDF of the $y$-map reconstructed from PR4 data using a harmonic-domain ILC (HILC), without localization in the pixel domain. The HILC $y$-map is clearly inferior to the NILC $y$-map with a much larger variance of the distribution. This highlights the importance of spatial localization to reconstruct thermal SZ signals, as NILC proceeds, while HILC is not a recommended component separation method for thermal SZ mapping. This aspect is further elaborated in Appendix \ref{app:hilc}

As can be seen from simulations in Fig.~\ref{fig:tSZ_pdf_plk_sims} (Appendix~\ref{app:sims}), which shows the contributions of various residual foregrounds to the 1-PDF of the NILC $y$-map, non-Gaussian foregrounds from Galactic emission and extragalactic infrared sources add reasonably low skewness to the positive tail of the $y$-map 1-PDF, while extragalactic radio sources add significant negative skewness to the distribution if not masked. Hence, we chose to use point-source masks from frequency channels below $217$\,GHz in order to suppress the negative contribution from radio sources in the PR4 and PR2 $y$-map 1-PDFs, while we kept infrared sources since masking them can cause some slight loss of power in the thermal SZ tail of the distribution. This choice of masking is justified as the infrared sources, which dominate at high frequencies, are not a major contaminant to the thermal SZ $y$-map (Fig.~\ref{fig:tSZ_pdf_plk_sims}). Furthermore, it is possible that a small number of these sources are actually unresolved SZ sources that have been mistakenly identified as infrared sources in catalogues. Point-source residuals and masking are further explored in Section~\ref{sec:residual_ps}.

To get a constraint on $\sigma_8$, we followed the approach presented by \citet{Wilson2012} and computed the unnormalized skewness $\langle \widehat{y}^{\,3}(p) \rangle$ of the PR4 NILC $y$-map using the same sky area and source mask as in \citet{planck2014-a28}. To get a rough estimate of the uncertainty, we also computed the skewness of the associated noise estimated from the half-difference of half-ring $y$-maps.
By applying the characteristic scaling relation of ${\langle \widehat{y}^{\,3}(p) \rangle\sim \sigma_8^{11}}$, we derived a value of $\sigma_8=0.76\pm 0.02$ for the PR4 NILC $y$-map, which is consistent with the value reported for the PR2 NILC y-map ($\sigma_8=0.77\pm 0.02$) in \citet{planck2014-a28}. It was anticipated, given the matching positive tails of the 1-PDF of the PR4 and PR2 NILC $y$-maps (Fig.~\ref{fig:tSZ_1PDF}).

\subsection{Angular power spectrum analysis} \label{sec:C_l}

The angular power spectrum of the thermal SZ effect has long been recognised as an important astrophysical and cosmological probe  \citep{Komatsu1999,Refregier2000,Komatsu2002}, as it integrates all the thermal SZ emission in the sky, both from diffuse, unbound hot gas and compact clusters of any mass and redshift. All-sky $y$-maps from the \textit{Planck} survey allow for computing the thermal SZ angular power spectrum over a large fraction of the sky and a relatively broad range of angular scales.

We use \texttt{NaMaster}\footnote{\url{https://namaster.readthedocs.io/en/latest}} \citep{2019MNRAS.484.4127A} to compute the angular power spectra of the $y$-maps. It allows us to reconstruct the thermal SZ power spectrum from the masked $y$-maps using a pseudo-\textit{$C_\ell$} estimation, while also taking care of the mode-coupling due to the application of the mask, the beam convolution, the pixelization, and the multipole binning. The PR2 and PR4 $y$-map power spectra are computed with the GAL-MASK and PS-MASK, leaving about $56$\% of the sky available after apodization of the PS-MASK. A custom binning scheme is defined in the plots with linear bins of width $\Delta\ell=3$ from multipoles $\ell=2$ to $30$ and logarithmic bins with $\Delta \log(\ell) = 0.05$ from $\ell=30$ onwards. This binning gives us $95$ band powers. 
All power spectra are computed up to $\ell=2048$ since the data at higher multipoles is consistent with noise due to the $10'$ resolution of the $y$-maps.

Once the point sources are masked, instrumental noise dominates the power at high multipoles. To correct for the noise bias in the estimated thermal SZ power spectrum, we compute the cross-power spectrum between the PR4 HR1 and HR2 $y$-maps since these two maps have mostly-uncorrelated noise due to half-ring data split (Section~\ref{sec:nilc_method}): 
\begin{equation} \label{eq:cl_noise_debiased}
    \widehat{C}_\ell^{\,\rm tSZ,~HR1\times HR2} = \frac{1}{2\ell+1}\sum_{\ell=2}^{\ell_{\rm max}} \widehat{y}_{\ell m}^{\,*~\rm HR1}\,\widehat{y}_{\ell m}^{\,\rm HR2},
\end{equation}
where $\widehat{y}_{\ell m}^{\,\rm HR1}$ and $\widehat{y}_{\ell m}^{\,\rm HR2}$ are the spherical harmonic coefficients of the PR4 HR1 and HR2 $y$-maps. The same is done for PR2 using public half-ring $y$-maps from \citet{planck2014-a28}.

\begin{figure}
	\includegraphics[width=\columnwidth]{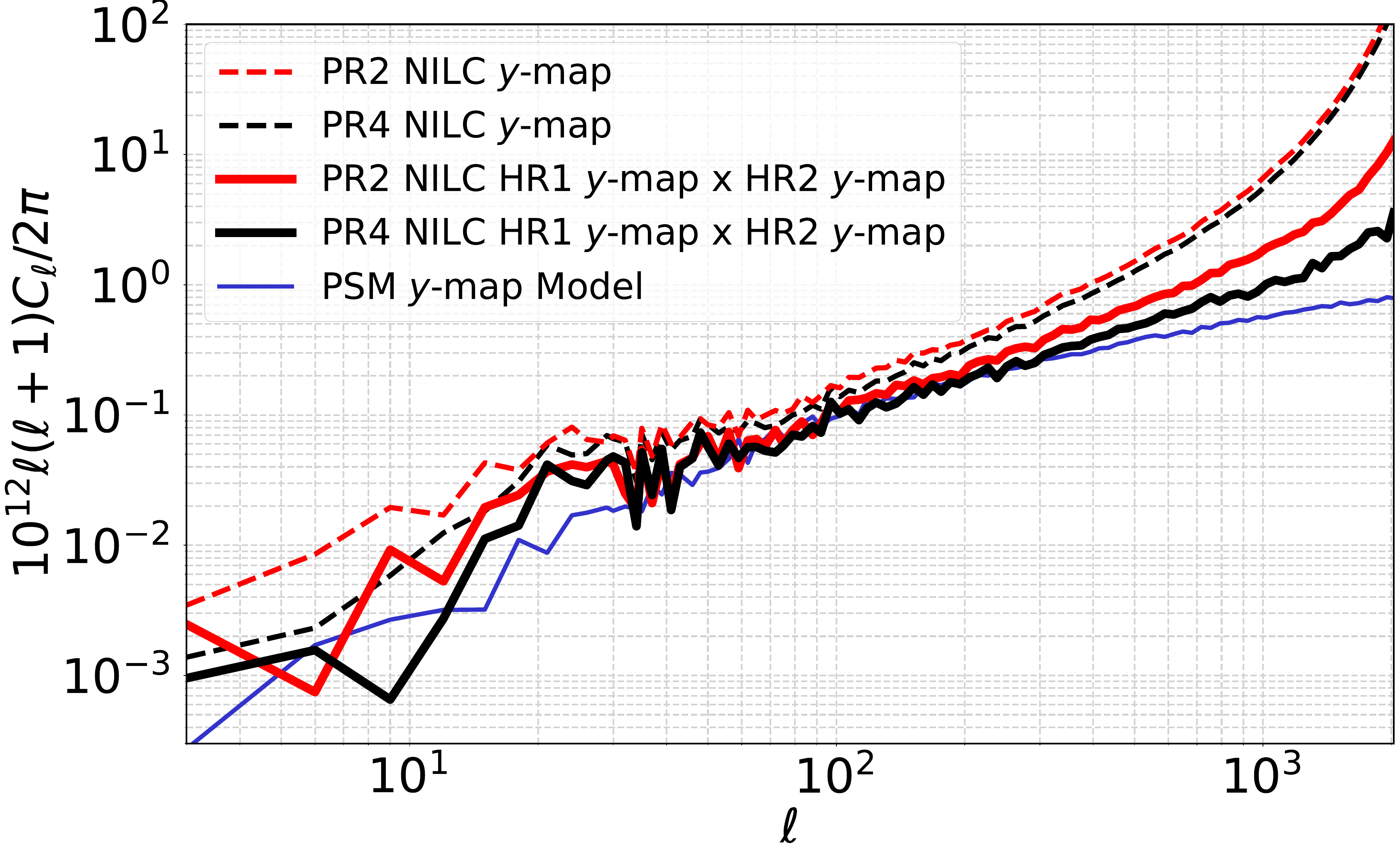}
    \caption{Thermal SZ angular power spectra from the PR2 NILC $y$-map (\emph{red}) and the PR4 NILC $y$-map (\emph{black}), before correction for the noise bias through the auto-power spectrum of the $y$-maps (\emph{dashed lines}) and after correction for the noise bias through the cross-power spectrum between the HR1 and HR2 $y$-maps (\emph{solid lines}). The PSM SZ model (\emph{thin blue line}) is shown as a reference.}
    \label{fig:tSZ_cl}
\end{figure}

Fig.~\ref{fig:tSZ_cl} shows the auto-power spectra of the PR2 and PR4 NILC $y$-maps (dashed lines) along with the reconstructed thermal SZ power spectra 
obtained from the cross-spectra between HR1 and HR2 $y$-maps (solid lines) for both PR2 (red) and PR4 (black). As a reference, the thermal SZ power spectrum from the Planck Sky Model \citep[PSM;][]{2013A&A...553A..96D}, which is used for the NILC analysis on \textit{Planck} simulations in Appendix \ref{app:sims}, is overplotted as a thin blue line in Fig.~\ref{fig:tSZ_cl}.

The power spectrum of the noise in the $y$-maps is also estimated using the half-difference of the half-ring $y$-maps, $(\text{HR1}\, y\text{-map} - \text{HR2}\, y\text{-map})/2$, and plotted in Fig.~\ref{fig:tSZ_noise_cl} for PR2 (red) and PR4 (black). 
The PR4 NILC $y$-map benefits from lower noise at all angular scales compared to the public PR2 NILC $y$-map, as a consequence of the reduced levels of noise in the \textit{Planck} PR4 data from the inclusion of 8\% more data (see Section~\ref{sec:residual_noise} for further discussion). As we will see, the reduced levels of noise in the PR4 data give the possibility to NILC to further minimize the variance of extragalactic foreground contamination (CIB and radio sources) at small angular scales.  

As we can see in Fig.~\ref{fig:tSZ_cl}, instrumental noise dominates the power at small angular scales (dashed lines), thus biasing the recovered thermal SZ power spectrum. After correcting for the noise bias using cross-power spectrum between half-ring $y$-maps (solid lines), there is still some remaining excess power at high multipoles in the reconstructed thermal SZ power spectrum which is associated mostly with residual CIB contamination in the $y$-maps, as also confirmed by the analysis on \textit{Planck} simulations in Fig.~\ref{fig:tSZ_cl_plk_sims} (see Appendix~\ref{app:sims}). In contrast, the excess power at low multipoles is due to residual Galactic foreground contamination at large angular scales. Clearly, the PR4 $y$-map power spectrum (solid black) shows much less power at low and high multipoles compared to the PR2 $y$-map power spectrum (solid red), which indicates lower residual foreground contamination in the PR4 NILC $y$-map (see Section~\ref{sec:residual_cib} for further discussion). The PR2 and PR4 $y$-map power spectra are more consistent at intermediate multipoles ($\ell \sim 30$-$300$) where the reconstructed thermal SZ signal dominates over the residual Galactic and extragalactic foreground contamination. Around the same range of multipoles ($\ell \sim 50$-$300$), the reconstructed signal is also close to the PSM model of thermal SZ emission. While the PR4 $y$-map conclusively has significantly lower contamination from extragalactic foregrounds at small angular scales compared to the PR2 $y$-map, the difference in power seen at low multipoles is not statistically significant because of the large cosmic variance expected from the non-Gaussian SZ signal \citep{Cooray2001,Komatsu2002,Bolliet2018}. 
Further analysis of the residual contamination in the thermal SZ $y$-maps is done in Section~\ref{sec:residuals}.

\section{Residual foreground and noise contamination in the PR4 $y$-map} \label{sec:residuals}

According to Fig.~\ref{fig:tSZ_cl_plk_sims} in Appendix~\ref{app:sims}, where the same NILC pipeline is applied to \textit{Planck} simulations, the major contaminants to the reconstructed $y$-map are the instrumental noise, the CIB and extragalactic radio sources at small angular scales and the Galactic foreground emission at large angular scales. In contrast, CMB and infrared sources are not major contaminants to the NILC $y$-map. While residual compact sources and Galactic foregrounds can be further mitigated by the application of appropriate masks, this is not possible for CIB whose emission is more diffuse and homogeneous over the sky.
In this section, we compare and quantify the residual contamination of the PR2 and PR4 NILC $y$-maps by the noise, the CIB and extragalactic compact sources.

\subsection{Noise} \label{sec:residual_noise}

The half-difference between the HR1 and HR2 PR4 $y$-maps gives us a noise map whose statistical properties are those of the actual residual noise fluctuations in the full (ring) PR4 $y$-map. Similarly, we get an estimate of the noise contamination in the PR2 $y$-map from the half-difference of the public PR2 half-ring NILC $y$-maps released by the \textit{Planck} Collaboration \citep{planck2014-a28}. The resulting $y$-noise maps are used to compare the noise characteristics of the PR4 and PR2 NILC $y$-maps.

\begin{figure}
	\includegraphics[width=\columnwidth]{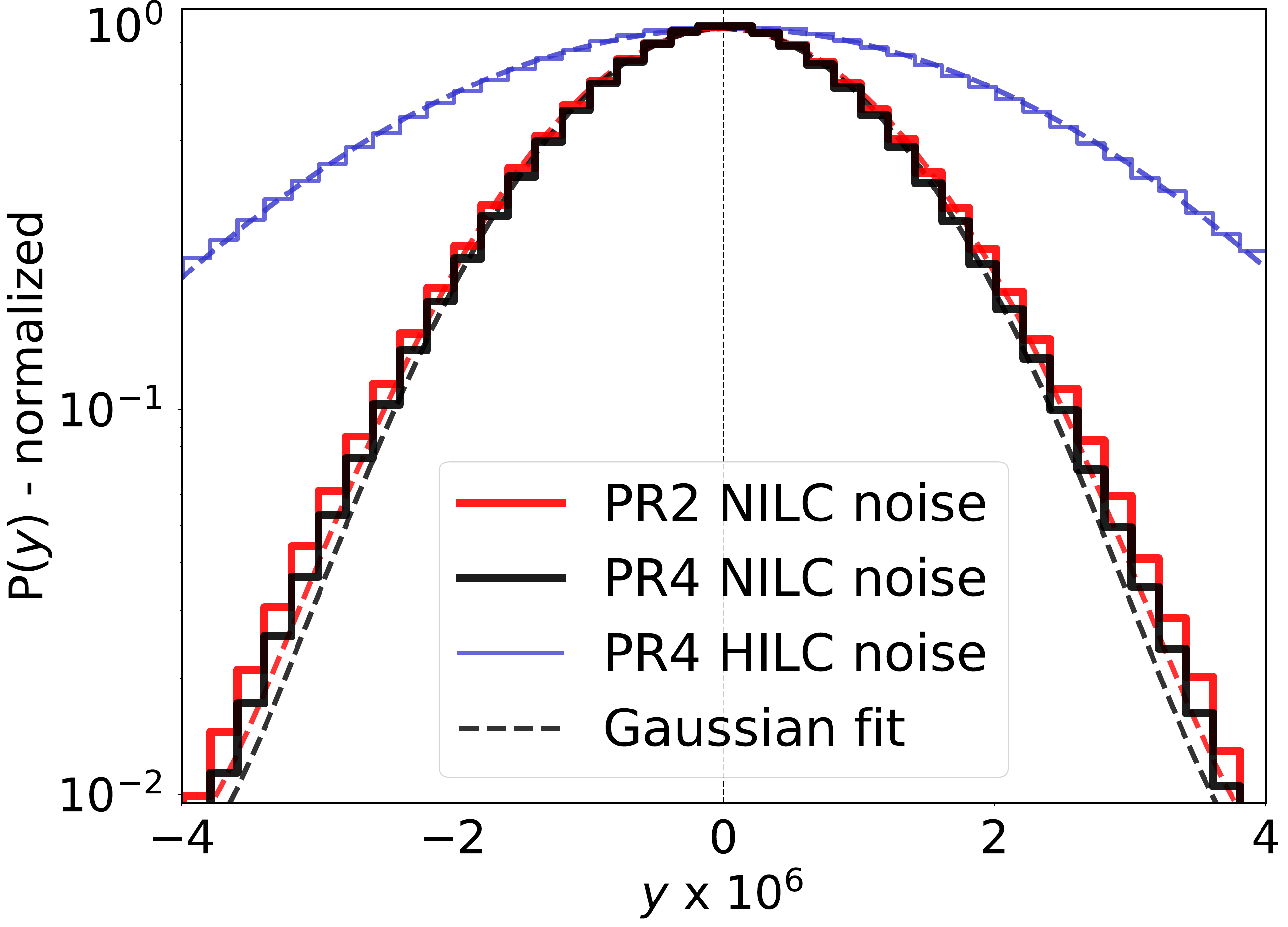}
     \caption{1-PDF of residual noise estimate from PR4 (\emph{black}) and PR2 (\emph{red}) NILC $y$-maps. Dashed lines show respective Gaussian fits. The noise 1-PDF from the PR4 HILC $y$-map (Appendix~\ref{app:hilc}) is also shown for comparison.}
    \label{fig:tSZ_noise_1PDF}
\end{figure}

Fig.~\ref{fig:tSZ_noise_1PDF} shows the binned normalized histogram (1-PDF) of the residual noise for the PR4 NILC $y$-map (black), the PR2 NILC $y$-map (red), and the HILC $y$-map (blue). The best-fitting Gaussian PDF (dashed line) is also shown in each case. As we can see, the noise distribution is mostly Gaussian in all $y$-maps.
The PR4 NILC $y$-map (best-fitting Gaussian noise standard deviation of $\sigma=1.12 \times 10^{-6}$) has lower noise as compared to the public PR2 NILC $y$-map (best-fit Gaussian noise standard deviation of $\sigma=1.16 \times 10^{-6}$), with a $6.8$\% reduction of the noise variance. The HILC $y$-map has comparatively much higher noise than any of the NILC $y$-maps, with a best-fit Gaussian standard deviation of $\sigma=2.27 \times 10^{-6}$. 

\begin{figure}
	\includegraphics[width=\columnwidth]{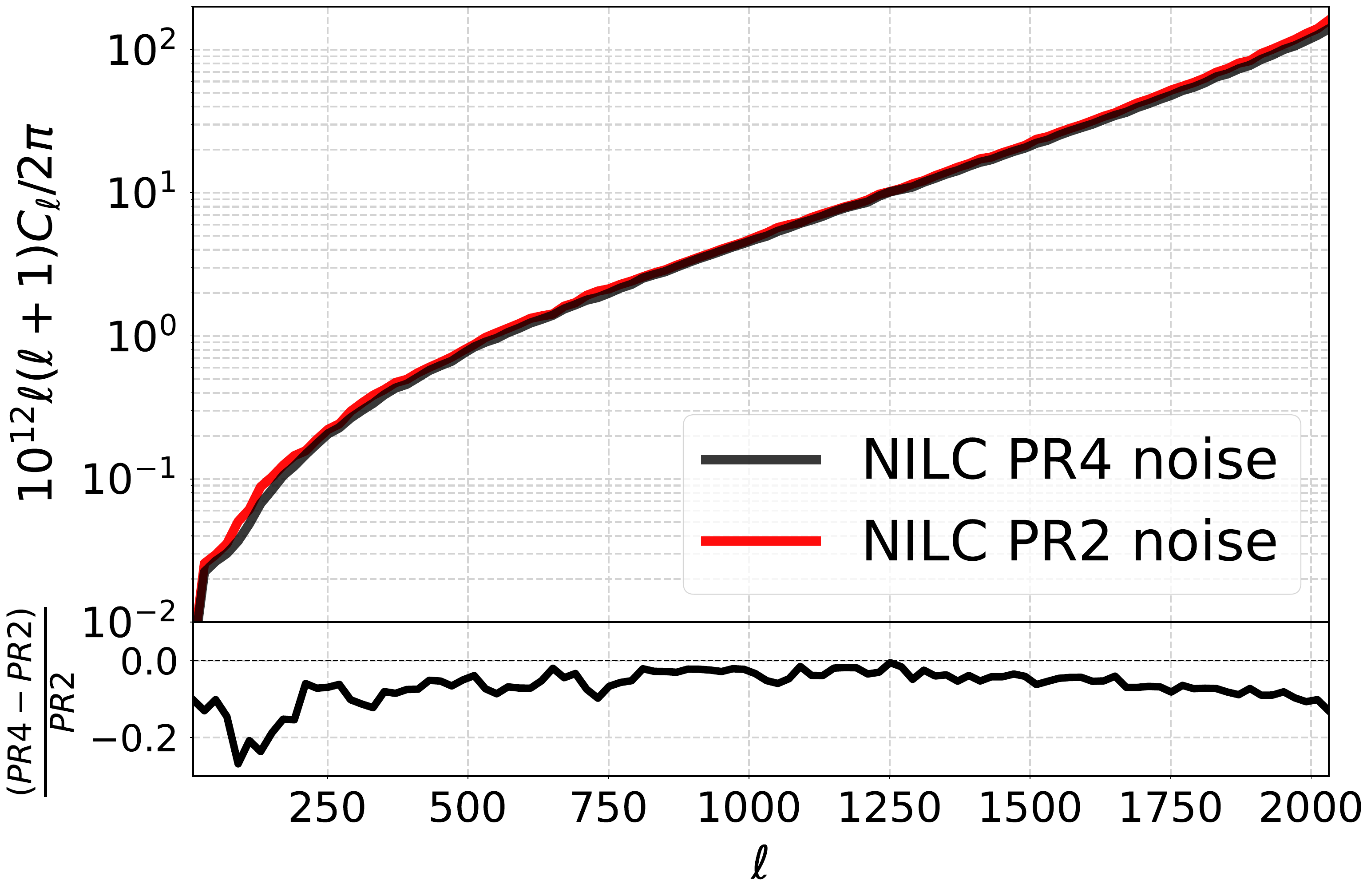}
    \caption{Angular power spectrum of residual noise in the PR4 and PR2 NILC $y$-maps (\emph{top}), and relative decrease of noise power in the PR4 $y$-map with respect to the PR2 $y$-map (\emph{bottom}).}
    \label{fig:tSZ_noise_cl}
\end{figure}

Fig.~\ref{fig:tSZ_noise_cl} shows the angular power spectrum of residual noise for the PR4 NILC $y$-map (black) and the PR2 NILC $y$-map (red) over $f_{\rm sky}=56$\% of the sky, as well as the relative decrease of noise power in the PR4 $y$-map with respect to the PR2 $y$-map across the multipoles, i.e. $\left(C_\ell^{\text{\,PR4\,} y\text{-noise}}-C_\ell^{\text{\,PR2\,} y\text{-noise}}\right)/C_\ell^{\text{\,PR2\,} y\text{-noise}}$. A linear binning of $\Delta\ell=20$ is used for this plot.
As evident from the bottom panel of Fig.~\ref{fig:tSZ_noise_cl}, the PR4 $y$-map has consistently lower noise power than the PR2 $y$-map at all angular scales. The mean percentage improvement of residual noise power over the multipole range $\ell=30$-$2048$ is 6.7\%, which is consistent with the result obtained from the 1-PDF. This improvement is due to the overall lower noise level in the \textit{Planck} PR4 data for the reasons listed in Section~\ref{sec:NPIPE_data}.

\subsection{Cosmic infrared background (CIB)} \label{sec:residual_cib}

The CIB is the most significant foreground contaminant to thermal SZ $y$-maps at small angular scales (see e.g. Fig.~\ref{fig:tSZ_cl_plk_sims} in Appendix~\ref{app:sims}). 
To assess and compare the levels of residual CIB contamination in the PR4 and PR2 $y$-maps, we compute the cross-power spectrum between these $y$-maps and \textit{Planck} CIB maps at $857$\,GHz. We choose $857$\,GHz as the frequency of the CIB templates because the $857$\,GHz channel of \textit{Planck} is not used at high multipoles $\ell>300$ for the reconstruction of the PR4 NILC $y$-map (see Table~\ref{tab:frequency_needlet_band}) and the PR2 NILC/MILCA $y$-maps \citep[see][]{planck2014-a28}. This enables the exclusion of spurious correlations between the noise of the $y$-maps and that of the $857$\,GHz CIB map, as the noise from different frequencies is uncorrelated, while the CIB is still highly correlated across frequencies \citep{planck2013-pip56}. Since the thermal SZ intensity is negligible at $857$\,GHz (about $3\%$ of its maximum intensity value at $353$\,GHz; see Fig.~\ref{fig:freq_response} and Table~\ref{tab:mixing_vector}), using CIB maps at $857$\,GHz also allows us to exclude spurious correlations that would be caused by residual SZ contamination in CIB templates of lower frequency.

\begin{figure}
	\includegraphics[width=\columnwidth]{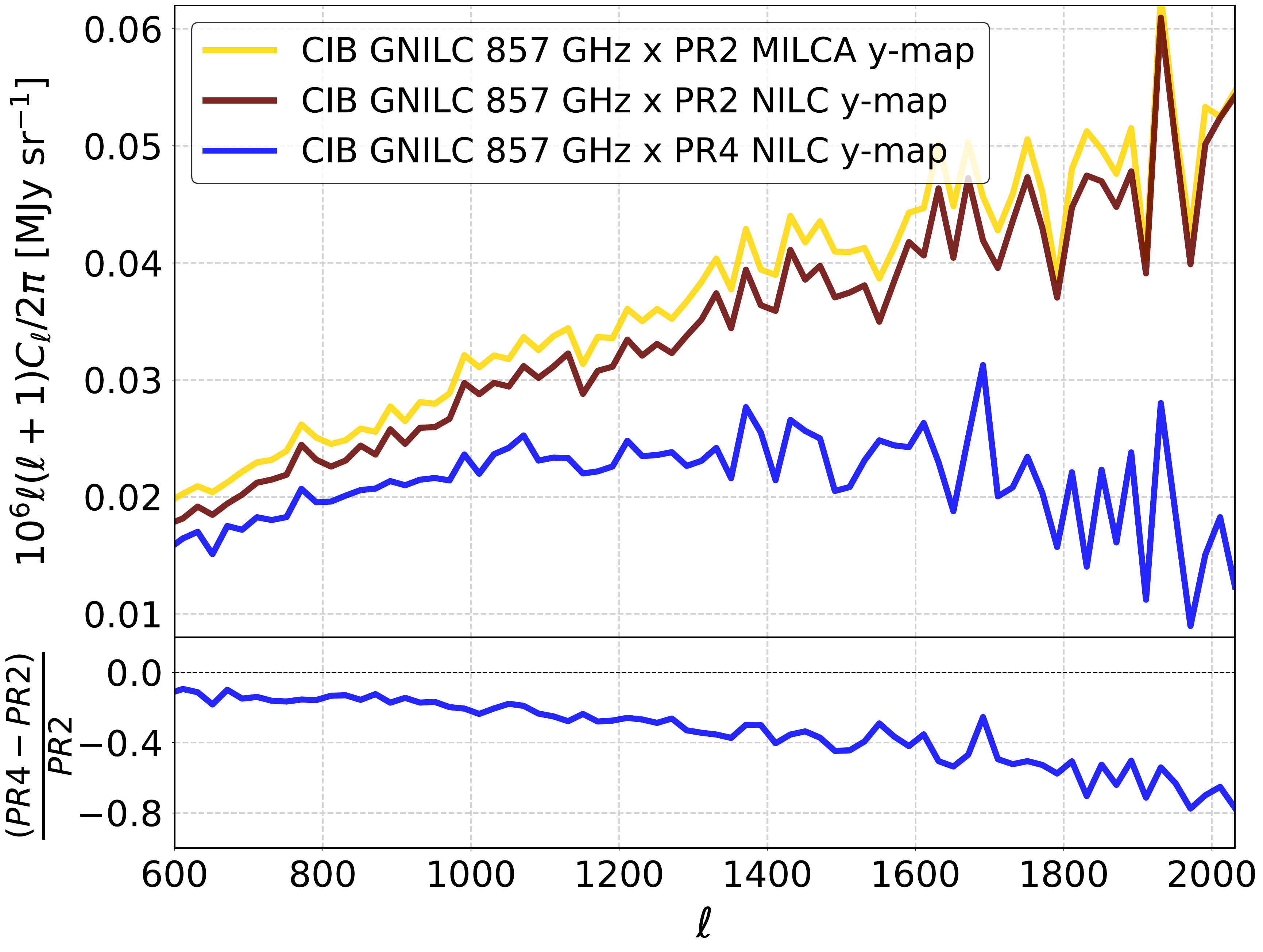}\\[0.2cm]
	\includegraphics[width=\columnwidth]{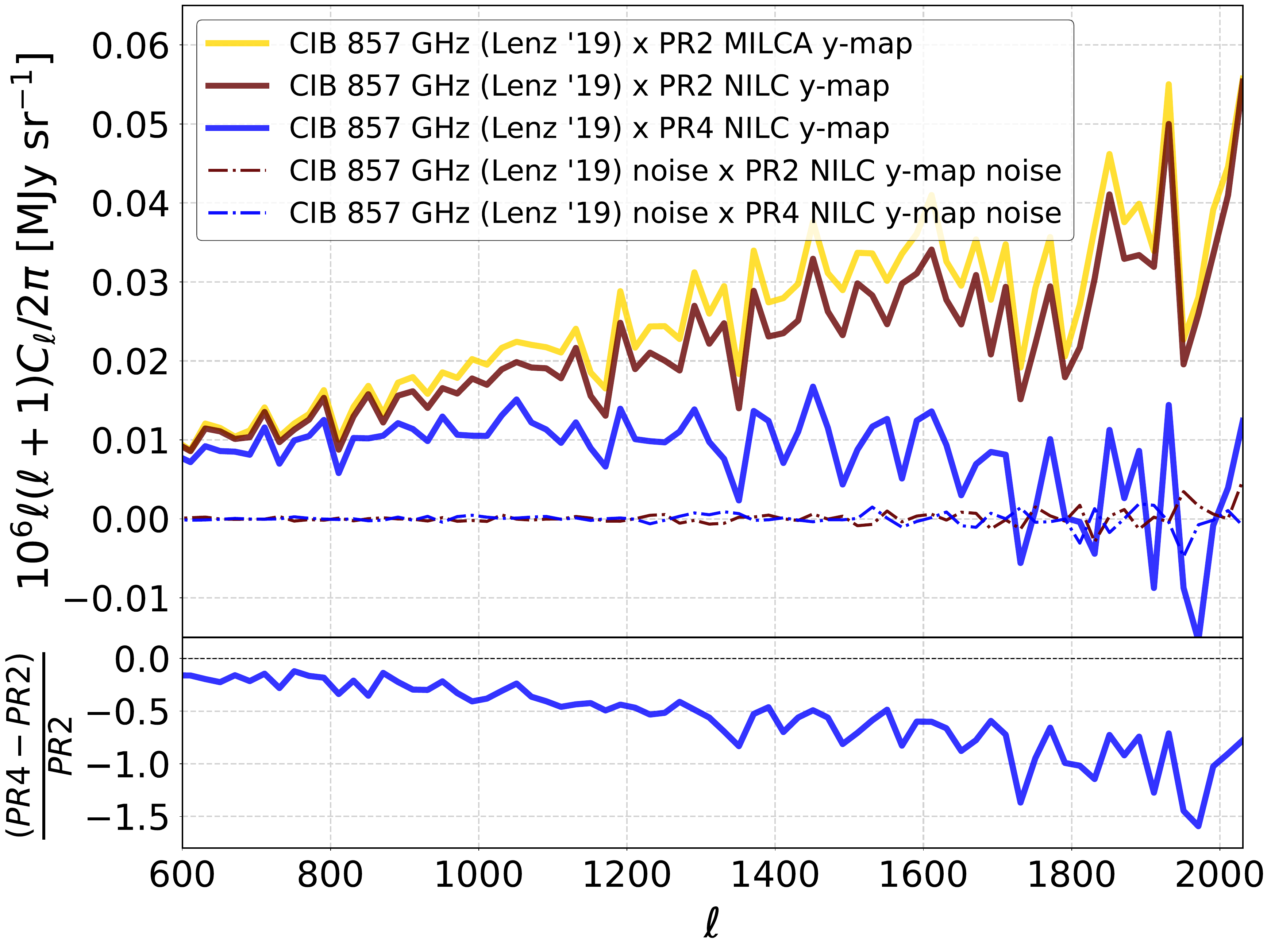}
    \caption{Levels of residual CIB contamination in the PR2 NILC / MILCA (\emph{red} / \emph{yellow}) and PR4  NILC (\emph{blue}) $y$-maps by taking cross spectra with \textit{Planck} CIB maps at $857$\,GHz. \emph{Top}: results with \textit{Planck} GNILC CIB map at $857$\,GHz ($f_{\rm sky}=50\%$). \emph{Bottom}: results with \citet{Lenz2019} CIB map at $857$\,GHz ($f_{\rm sky}=15\%$). The cross-spectrum between CIB noise and $y$-map noise estimates is also shown (dashed-dotted lines). 
    }
    \label{fig:tSZ_cib_GNILC}
\end{figure}

We use two independent CIB templates for our analysis (see Section~\ref{sec:foreground_tracers}): the \textit{Planck} GNILC CIB map at $857$\,GHz \citep{planck2016-XLVIII} covering $57$\% of the sky and the \textit{Planck}-based CIB map at $857$\,GHz from \citet{Lenz2019} which covers $18$\% of the sky. Combining with the GAL-MASK and PS-MASK of the $y$-maps, the CIB cross $y$-map power spectrum is computed with \texttt{NaMaster} over $50$\% of the sky when using the \textit{Planck} GNILC CIB map and over $15$\% of the sky when using the \citet{Lenz2019} CIB map. The results are shown in Fig.~\ref{fig:tSZ_cib_GNILC} for high multipoles $\ell > 600$, with a linear binning of $\Delta\ell=20$.
For either CIB template, a consistent pattern emerges, showing a much stronger correlation with the PR2 $y$-maps (red / yellow) than with the new PR4 $y$-map (blue).  The residual CIB contamination in the PR4 NILC $y$-map is thus considerably lesser than the CIB contamination of the public PR2 $y$-maps. This contributes to lesser the overall contamination of the $y$-map power spectrum at high multipoles as observed in Fig.~\ref{fig:tSZ_cl}.

The lower sub-panels of Fig.~\ref{fig:tSZ_cib_GNILC} display the relative decrease of CIB contamination in the PR4 NILC $y$-map with respect to the PR2 NILC $y$-map over the multipoles, i.e. $\left(C_\ell^{\,\text{PR4}\,y\, \times\, \text{CIB}} - C_\ell^{\,\text{PR2}\,y\, \times\, \text{CIB}}\right)/C_\ell^{\,\text{PR2}\,y\, \times\, \text{CIB}}$. By averaging over the multipole range $\ell=600$-$2048$, we infer a $34.2$\% decrease in residual CIB power in the PR4 NILC $y$-map compared to the PR2 NILC $y$-map over $50$\% of the sky when using the \textit{Planck} GNILC CIB template (top panel), and a $56.7$\% decrease over $15$\% of the sky when using the \citet{Lenz2019} CIB template (bottom panel).
As a matter of fact, the lower noise variance in PR4 data compared to PR2 data allows the NILC pipeline to further minimize the variance of the foregrounds at high multipoles, such as CIB (Fig.~\ref{fig:tSZ_cib_GNILC}) and extragalactic compact sources (Section~\ref{sec:residual_ps}). Since the variance of instrumental noise is much larger than the variance of CIB fluctuations in the \textit{Planck} data, even a few per cent reduction of noise in the PR4 data can make a big difference in the CIB variance minimization by NILC for the $y$-map.

CIB maps reconstructed from half-ring data splits being available for the \citet{Lenz2019} template, we obtained a CIB noise map from the half-difference of half-ring CIB maps which we cross-correlated with the $y$-noise maps also obtained from the half-ring data split. As already anticipated and shown in the bottom panel of Fig.~\ref{fig:tSZ_cib_GNILC} (dashed-dotted lines), the noise in the $857$\,GHz CIB map is mainly uncorrelated with the noise in the $y$-maps due to the \textit{Planck} 857 GHz channel not being used at high multipoles for the construction of the $y$-maps. Hence, the noise contribution to the cross-spectrum analysis is mostly negligible and cannot account for the difference seen in Fig.~\ref{fig:tSZ_cib_GNILC}.
 
We might wonder about the possibility of a bias in the cross-spectra comparison due to the use of GNILC CIB maps from the \textit{Planck} PR2 release, which is the same data used for one of the $y$-maps. However, given that we used the $857$\,GHz GNILC CIB map and the \textit{Planck} $857$\,GHz map was not used for CIB-relevant multipoles in either PR2 or PR4 $y$-maps reduces the likelihood of such a bias. Additionally, our findings are supported by the fact that the \citet{Lenz2019} CIB map, based not on PR2 or PR4 but on \textit{Planck} PR3 data, yields similar cross-spectra results.

\subsection{Extragalactic compact sources} \label{sec:residual_ps}

Radio and infrared (IR) compact sources are another important foreground contaminant to thermal SZ emission at small angular scales. They are usually handled by the use of point-source masks. Radio sources from active galactic nuclei (AGNs) emitting synchrotron radiation are mostly detected at low frequencies below $217$\,GHz, while IR sources from star-forming dusty galaxies are detected at high frequencies \citep{planck2014-a35}.
 Hence, by masking the point sources detected in either low or high \textit{Planck} frequency channels, we can explore the residual contamination from radio and IR sources separately and determine the optimal masking strategy. 
 
\begin{figure}
    \includegraphics[width=\columnwidth]{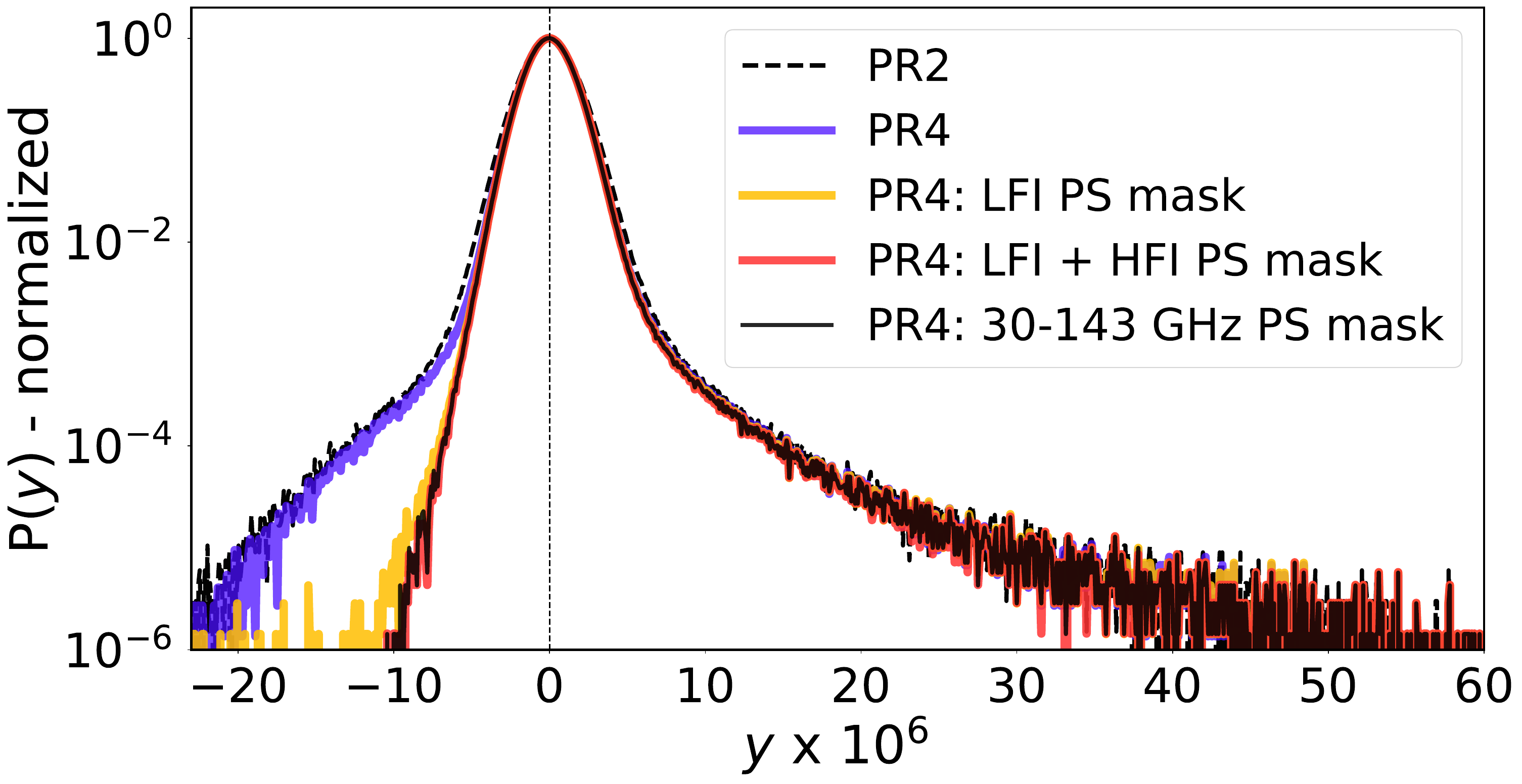}
	\caption{1-PDF of PR4 NILC $y$-map without point-source mask (\emph{purple}), with LFI source masks (\emph{yellow}), with $30$ to $143$\,GHz source masks (\emph{black}), and with all LFI and HFI source masks (\emph{red}). The major source contamination comes from radio sources in LFI channels. The PR4 NILC $y$-map (\emph{purple}) has a lower level of radio-source contamination than the PR2 NILC $y$-map (\emph{dashed black}).}
    \label{fig:tSZ_pdf_point_sources}
\end{figure}

 In Fig.~\ref{fig:tSZ_pdf_point_sources}, we explore the impact of radio and IR source contamination on the 1-PDF of the PR4 and PR2 $y$-maps by masking them with three different point-source masks: 
\begin{enumerate}
    \item A point-source mask defined as the union of point-source masks from all \textit{Planck} channels, which masks both IR and radio sources (red line).
    \item A point-source mask defined as the union of the point-source masks from the three \textit{Planck} LFI channels ($30$, $44$ and $70$\,GHz), which masks only radio sources (yellow line).
    \item A point-source mask defined as the union of the point-source masks from \textit{Planck}'s low frequencies below $217$\,GHz, where the spectral response of the thermal SZ effect is negative. This means that all the point sources that end up as negative spots in the $y$-map are masked (black line).
\end{enumerate}

 The NILC weights give unit response to the SED of the thermal SZ signal across \textit{Planck} channels. As the frequency dependence of the thermal SZ effect is negative at frequencies below 217 GHz (Fig.~\ref{fig:freq_response}), and the majority of the radio sources are detected in these frequencies, they end up as negative point sources in the reconstructed Compton $y$-maps. The opposite is true for IR sources in general. As the distribution of extragalactic sources over the sky is also non-Gaussian, radio sources contribute as a negatively skewed tail in the 1-PDF of the reconstructed $y$-map. In contrast, IR sources contribute to the positively skewed tail of the distribution (see Fig.~\ref{fig:tSZ_pdf_plk_sims} in Appendix~\ref{app:sims} for evidence from simulations).
 This assertion is supported by Fig.~\ref{fig:tSZ_pdf_point_sources}, where the drop of the negative non-Gaussian tail in the 1-PDF of the PR4 $y$-map is largely obtained by masking only radio sources detected in \textit{Planck} LFI channels (yellow line versus purple line).

Masking all detected \textit{Planck} sources in the $y$-map (red line) does not significantly attenuate the negative tail of the distribution beyond what is achieved by masking only the sources detected at frequencies $<217$\,GHz (black line). These two cases also show only marginal improvements compared to masking only LFI radio sources (yellow line). This is evident from the variance of the \mbox{1-PDF}, which is reduced by $5\%$ when masking sources detected below $217$\,GHz (black line) compared to the unmasked distribution (purple line), while a further reduction of only $0.5\%$ is achieved by masking all sources (red line). 
In addition, when all sources in the $y$-map are masked (red line), there is only a negligible drop compared to the black line in the positive non-Gaussian tail characteristic of thermal SZ (skewness reduced by only $1\%$). 
This suggests that masking sources detected at frequencies $\geq 217$\,GHz have negligible benefits, while there is also a risk of losing part of the non-Gaussian SZ signal due to erroneous masking of unresolved compact thermal SZ sources mistaken as IR sources. Therefore, the choice is made to mask out only sources below $217$\,GHz in the $y$-map using the corresponding \textit{Planck} compact source masks (black line). This approach allows for a higher sky fraction without significant contamination of the signal.

As visible from Fig.~\ref{fig:tSZ_pdf_point_sources} before masking, the negative non-Gaussian tail in the 1-PDF is slightly larger for the PR2 $y$-map (dashed black line) compared to the PR4 $y$-map (purple line), which suggests lower residual contamination from radio sources in the PR4 $y$-map. Again, the lower noise levels in the PR4 data allow the NILC pipeline to further reduce the extragalactic foreground contamination from CIB and radio sources in the $y$-map.

%%%%%%%%%%%%%%%%%%%%%%%%%%%%%%%%%%%%%%%%%%%%%%%%%%%%%%%%%%%%%%%%%%%%%%%%%%%%%%%%%%%%%%%%%%%%%%%%%%%%

\section{Conclusion} \label{sec:conclusions}

A new thermal SZ Compton $y$-map has been produced over 98\% of the sky by implementing a tailored NILC pipeline on the nine \textit{Planck} frequency maps ranging from $30$ to $857$\,GHz, as provided by the \textit{Planck}  PR4 data release. The newly introduced $y$-map represents a substantial improvement over the \textit{Planck} PR2 $y$-maps that were previously released by the \emph{Planck} Collaboration \citep{planck2014-a28}. The \textit{Planck} PR4 data feature reduced levels of noise and systematics, which translates into lower levels of noise and foreground contamination in the new PR4 NILC $y$-map, compared to the public \textit{Planck} PR2 $y$-maps.

Several tests have been conducted with map inspections, one-point and two-point statistics to validate the quality of the PR4 NILC $y$-map. These tests reveal that the noise has been reduced by about 7\% in the PR4 NILC $y$-map, while the residual contamination from CIB has decreased by more than 34\% compared to the PR2 $y$-maps. Moreover, the PR4 NILC $y$-map exhibits lower levels of large-scale striping from residual $1/ f$ noise compared to the public \textit{Planck} PR2 $y$-maps and reduced contamination from extragalactic radio sources. 
The constraint on the cosmological parameter ${\sigma_8 = 0.76\pm 0.02}$, obtained from the skewness analysis of the PR4 NILC $y$-map, remains consistent with the constraint derived from the PR2 NILC $y$-map analysis as reported in \citet{planck2014-a28}.

The \textit{Planck} PR4 NILC $y$-map, as well as  the associated half-ring $y$-maps and masks, are publicly available at \url{https://doi.org/10.5281/zenodo.7940376} and in the \href{https://pla.esac.esa.int/}{Planck Legacy Archive}.

We are considering several extensions to the current analysis for future studies. One approach is to incorporate external full-sky data in the NILC pipeline, along with the \textit{Planck} PR4 channel maps, to enhance foreground cleaning in the Compton $y$-map. This idea, similar to that of \citet{Kusiak2023} for CMB, uses additional channels from external data that trace foreground emission.

Another extension is to release a "CIB-free" PR4 $y$-map for cross-correlation studies with large-scale structure (LSS) tracers such as lensing maps. To achieve this, constrained ILC methods \citep*{2011MNRAS.410.2481R,Remazeilles2021} can be used to \emph{deproject} the spectral moments of the CIB \citep*{Chluba2017} from the PR4 data. Such an approach may significantly reduce biases caused by residual CIB-LSS correlations in thermal SZ-LSS cross-correlation studies. Recently, deprojection of CIB moments was applied to cluster detection algorithms and demonstrated favourable outcomes in simulations \citep*{2023MNRAS.522.5123Z}. Although the release of a CIB-free Compton $y$-map is planned, it is expected to have higher overall noise variance compared to the \emph{Planck} PR4 NILC $y$-map due to the additional CIB constraints imposed on NILC. With greatly reduced CIB contamination and minimal overall variance, the PR4 NILC $y$-map may already provide a reliable thermal SZ template for cross-correlation studies.

The last proposed extension is to account for relativistic corrections to the thermal SZ SED in the NILC pipeline by incorporating the average temperature of the \emph{Planck} clusters, as suggested by \citet{Remazeilles2019}. This extension is warranted for the \emph{Planck} PR4 $y$-map, because even though the relativistic SZ correction is faint, it is still expected to contribute statistically to the signal at \emph{Planck} sensitivity.

%%%%%%%%%%%%%%%%%%%%%%%%%%%%%%%%%%%%%%%%%%%%%%%%%%%%%%%%%%%%%%%%%%%%%%%%%%%%%%%%%%%%%%%%%%%%%%%%%%%%

\section*{Acknowledgements}

We thank Marcos L\'opez-Caniego for providing point-source masks for PR4 data.
We also thank Reijo Keskitalo and Hans Kristian Eriksen for valuable advice and comments on the PR4 (NPIPE) data release. The authors would like to thank the Spanish Agencia Estatal de Investigaci\'on (AEI, MICIU) for the financial support provided under the project PID2019-110610RB-C21.
JC also acknowledges financial support from the \textit{Concepci\'on Arenal Programme} of the Universidad de Cantabria.
MR also acknowledges financial support from the CSIC programme 'Ayuda a la Incorporaci\'on de Cient\'ificos Titulares' provided under the project 202250I159. Some of the
presented results are based on observations obtained with Planck,\footnote{http://www.esa.int/Planck} an ESA science mission
with instruments and contributions directly funded by ESA Member States, NASA, and
Canada.
This work made use of \texttt{Python} packages like \texttt{pymaster} \citep{2019MNRAS.484.4127A}, \texttt{astropy} \citep{2022ApJ...935..167A}, \texttt{scipy} \citep{2020SciPy-NMeth}, and \texttt{matplotlib} \citep{Hunter:2007}. Some of the results in this paper have been derived using the \texttt{healpy} \citep{2019JOSS....4.1298Z} and \textit{HEALPix} \citep{2005ApJ...622..759G} packages. 
We acknowledge the use of the PSM, developed by the Component Separation Working Group (WG2) of the \textit{Planck} Collaboration.

%%%%%%%%%%%%%%%%%%%%%%%%%%%%%%%%%%%%%%%%%%%%%%%%%%
\section*{Data Availability}

The PR4 thermal SZ data produced as a result of this study is available in Zenodo, at \url{https://doi.org/10.5281/zenodo.7940376} and in the \href{https://pla.esac.esa.int/}{Planck Legacy Archive}. The input \textit{Planck} PR4 data, the PR2 $y$-maps, the Galactic mask, and the GNILC CIB maps were taken from the Planck Legacy Archive.
The \citet*{Lenz2019} CIB maps are available at \url{https://doi.org/10.7910/DVN/8A1SR3}. The IRAS 100-micron map is available at \url{https://www.ipac.caltech.edu/doi/irsa/10.26131/IRSA94}.

%%%%%%%%%%%%%%%%%%%% REFERENCES %%%%%%%%%%%%%%%%%%

\bibliographystyle{mnras}
\bibliography{references,Planck_bib}

%%%%%%%%%%%%%%%%%%%%%%%%%%%%%%%%%%%%%%%%%%%%%%%%%%

%%%%%%%%%%%%%%%%% APPENDICES %%%%%%%%%%%%%%%%%%%%%

\appendix

\section{\textit{Planck} Simulations} \label{app:sims}

To get a sense of the behaviour of the different foreground residuals in the thermal SZ $y$-map, but also to validate our pipeline, we applied the same NILC algorithm to simulated data having the characteristics of the \textit{Planck} data.

The Planck Sky Model \citep[PSM;][]{2013A&A...553A..96D} is used to produce a set of sky maps at \textit{Planck} frequencies with contributions from thermal SZ and kinetic SZ effects, CMB, CIB, radio and infrared sources, Galactic foregrounds (thermal dust, synchrotron, free-free, anomalous microwave emission) and instrumental noise. The models used to simulate each of these components in the \textit{Planck} frequency bands are described in detail in \citet{2013A&A...553A..96D}. The assumed cosmological parameter values are those from the \textit{Planck} 2018 best-fitting model \citep{planck2016-l06}. Unlike PR4 data, the simulated data have symmetric Gaussian beams and monochromatic bandpasses and do not include $1/f$ noise but homogeneous Gaussian white noise with \textit{Planck} channel sensitivities.

The simulated \textit{Planck} frequency maps went through the same NILC pipeline as described in Section~\ref{sec:nilc_method} to derive NILC weights and reconstruct the thermal $y$-map. Each individual foreground and noise component of the simulation is propagated with the same NILC weights to reconstruct their respective residual map. 
Characterizing the residual contamination of the reconstructed $y$-map is then possible by computing the 1-PDF and the angular power spectra of the residual map of each contaminant.

\begin{figure}
	\includegraphics[width=\columnwidth]{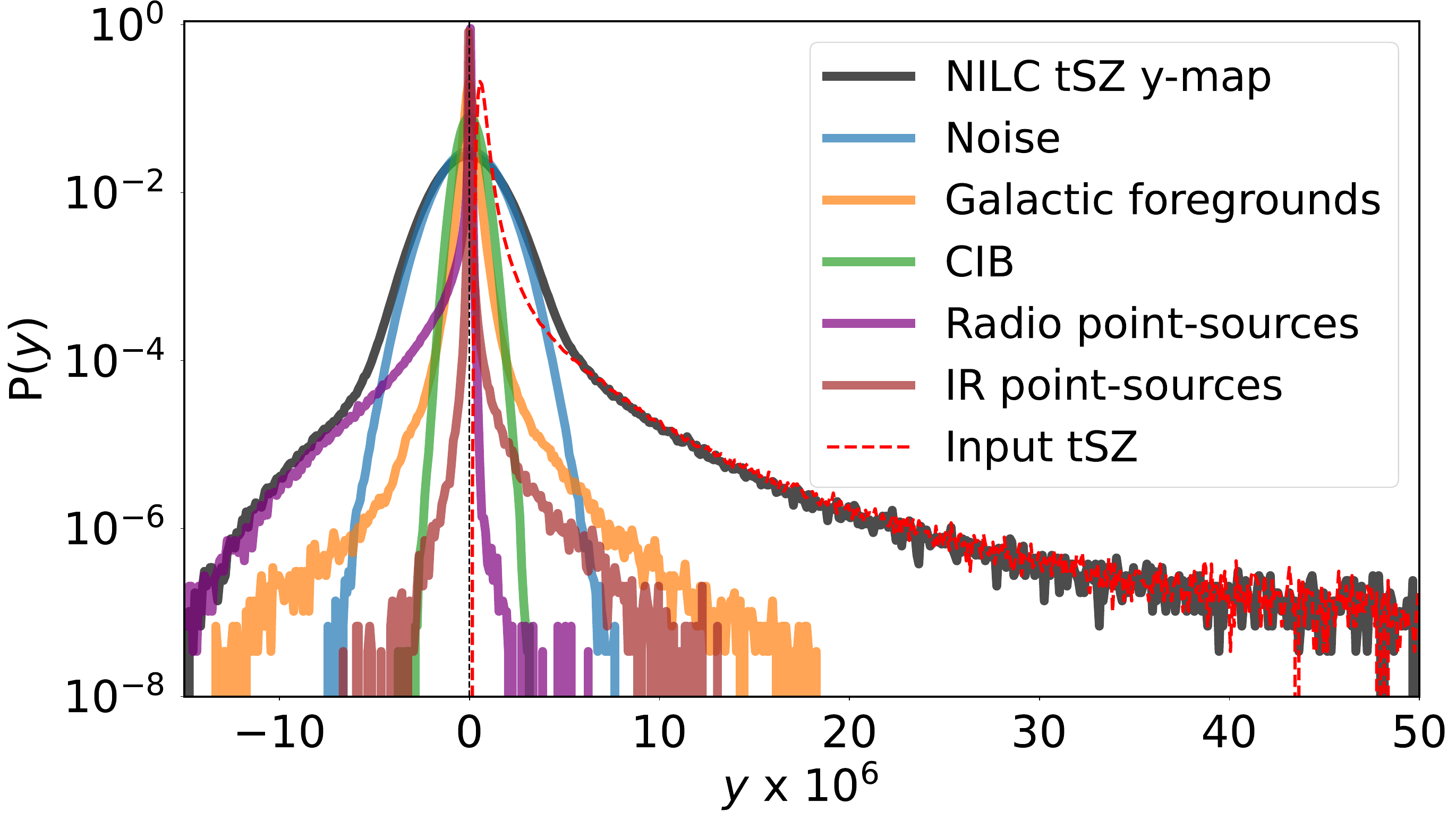}
     \caption{1-PDF of  NILC $y$-map (\emph{solid black}) versus input $y$-map (\emph{dashed red}), and individual contributions from various residual foregrounds  (\emph{coloured solid lines}) for \textit{Planck} PSM simulations.}
    \label{fig:tSZ_pdf_plk_sims}
\end{figure}

\begin{figure}
	\includegraphics[width=\columnwidth]{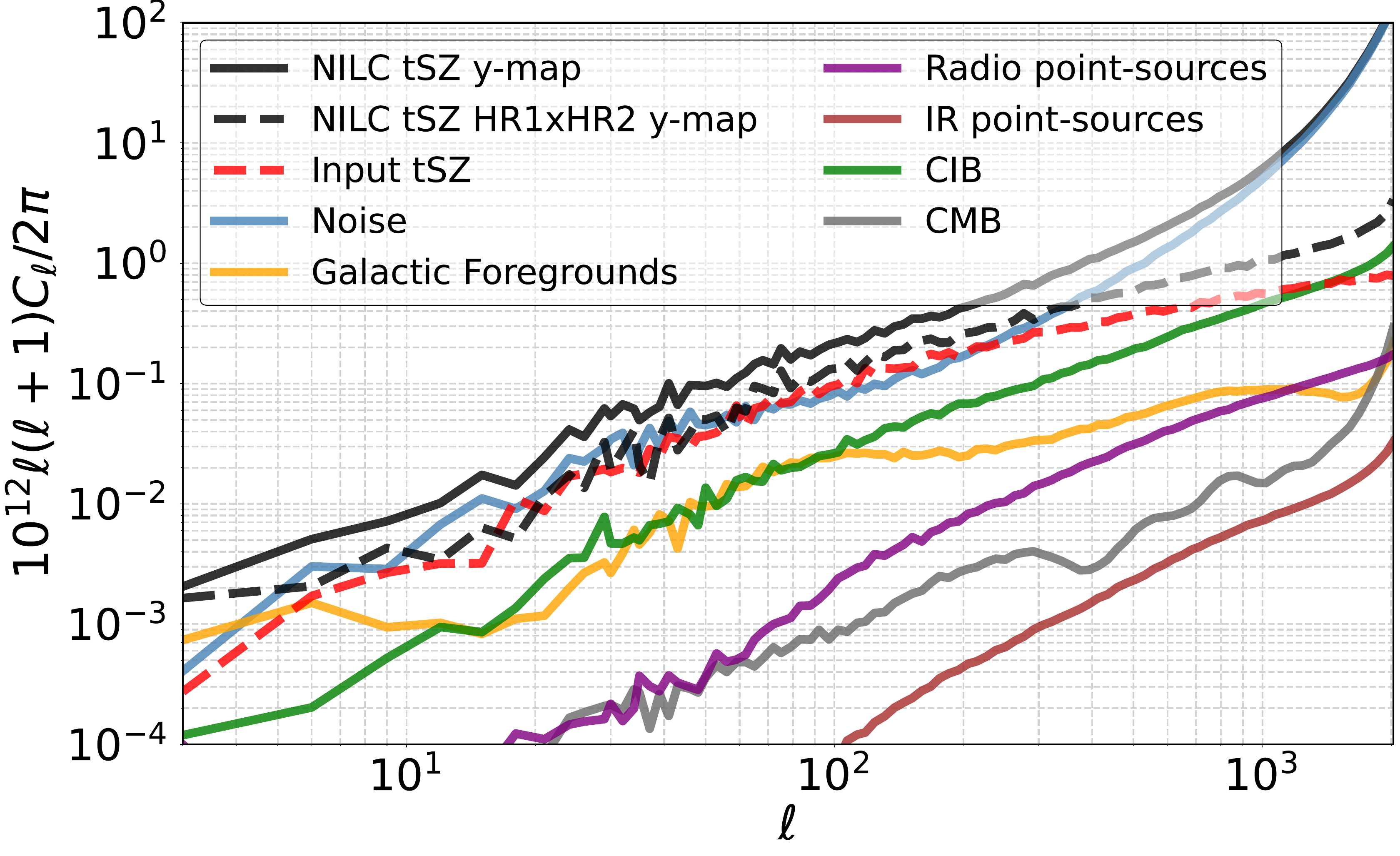}
   \caption{NILC thermal SZ and residual foreground power spectra for \textit{Planck} PSM simulations.}
    \label{fig:tSZ_cl_plk_sims}
\end{figure}

Fig.~\ref{fig:tSZ_pdf_plk_sims} shows the 1-PDF of the simulated input thermal SZ signal (dashed red), as well as the 1-PDF of the thermal SZ signal reconstructed from NILC (solid black) and that of each residual contaminant (solid coloured lines). 
Only pixels outside the Galactic plane mask (GAL-MASK) have been considered. Point-sources were not masked because we want to assess the contribution of radio and IR sources separately. 
In agreement with what we saw in Fig.~\ref{fig:tSZ_pdf_point_sources} for \textit{Planck} PR4 data (Section~\ref{sec:residual_ps}), the radio sources (purple line in Fig.~\ref{fig:tSZ_pdf_plk_sims}) contribute dominantly in the negative tail and sparingly to the positive region. This is because radio sources are significant at frequencies below $143$\,GHz, where the frequency response of the thermal SZ is negative. Hence, they end up as negative residual sources when propagated through NILC.
The opposite is largely true for the IR source residuals (brown line); they contribute primarily to the positive tail of the distribution and sub-dominantly to the negative tail. Nevertheless, IR source residuals are low enough so that the characteristic positive non-Gaussian tail of the thermal SZ \citep{Rubino-Martin2003} is almost perfectly recovered as can be seen by the overlapping positive tails of the input and output thermal SZ PDFs. This supports our findings in Section~\ref{sec:residual_ps} and our strategy to mask only radio sources detected in the frequency range where thermal SZ has a negative frequency response.

Residual Galactic foregrounds (orange line in Fig.~\ref{fig:tSZ_pdf_plk_sims}) contribute equally to the negative and positive non-Gaussian tails of the 1-PDF of the reconstructed $y$-map. Finally, excess variance arises mainly from residual noise (blue), CIB (green) and Galactic foregrounds (orange).

Fig.~\ref{fig:tSZ_cl_plk_sims} shows the angular power spectra of the input thermal SZ (dashed red), the output NILC thermal SZ obtained from auto-spectra (solid black) and from cross-spectra (i.e., free from noise bias; dashed black), the noise (blue), and the residual foreground components. Consistently with the PR4 data analysis, we apply the GAL-MASK and the PS-MASK\footnote{The use of the PS-MASK is appropriate here because the simulated point sources in the PSM rely on existing catalogues.} to compute the power spectra. As can be seen, all foregrounds are mitigated below the thermal SZ signal by NILC over a wide range of multipoles. The CIB (green) and the noise (blue) are the main residual contaminants at small angular scales (high multipoles) after point-source masking, while Galactic foregrounds (orange) add residual excess power mostly at large angular scales. Residual contamination from CMB (grey) is sub-dominant at all angular scales in the NILC $y$-map.

\section{Harmonic ILC for thermal SZ} \label{app:hilc}

For the sake of comparison, we also implemented a Harmonic Internal Linear Combination (HILC) on the PR4 data. The HILC operates in a spherical harmonic domain with perfect localization but lacks spatial localization in the pixel domain in contrast to NILC.
Even though the HILC may perform well in extracting homogeneous and isotropic signals like the CMB \citep*[e.g.][]{Tegmark2003}, it is definitely not the best option to reconstruct localized signals like the thermal SZ, as we show here. 

\begin{figure}
	\includegraphics[width=\columnwidth]{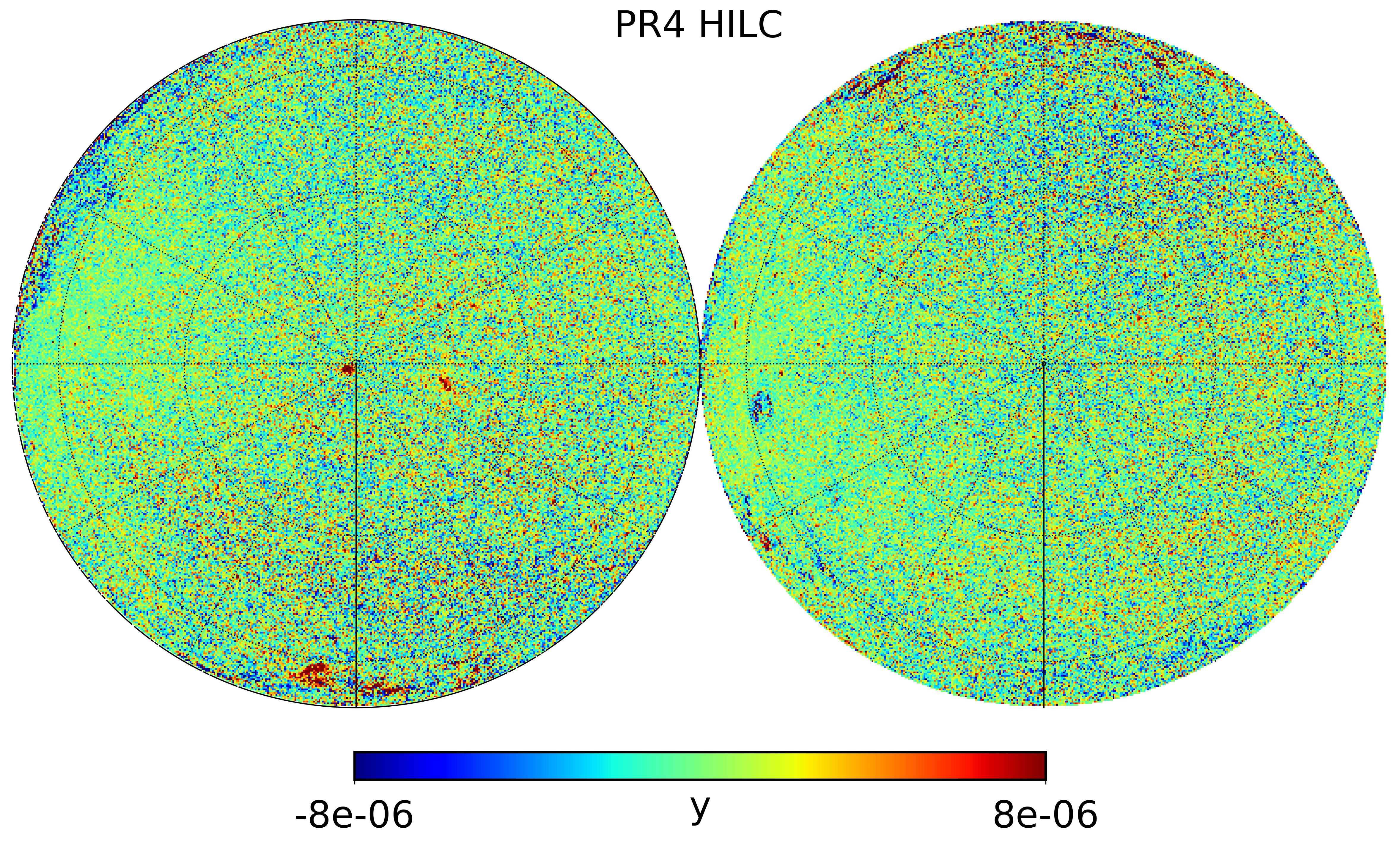}
 
    \vspace{0.5cm}
    
	\includegraphics[width=\columnwidth]{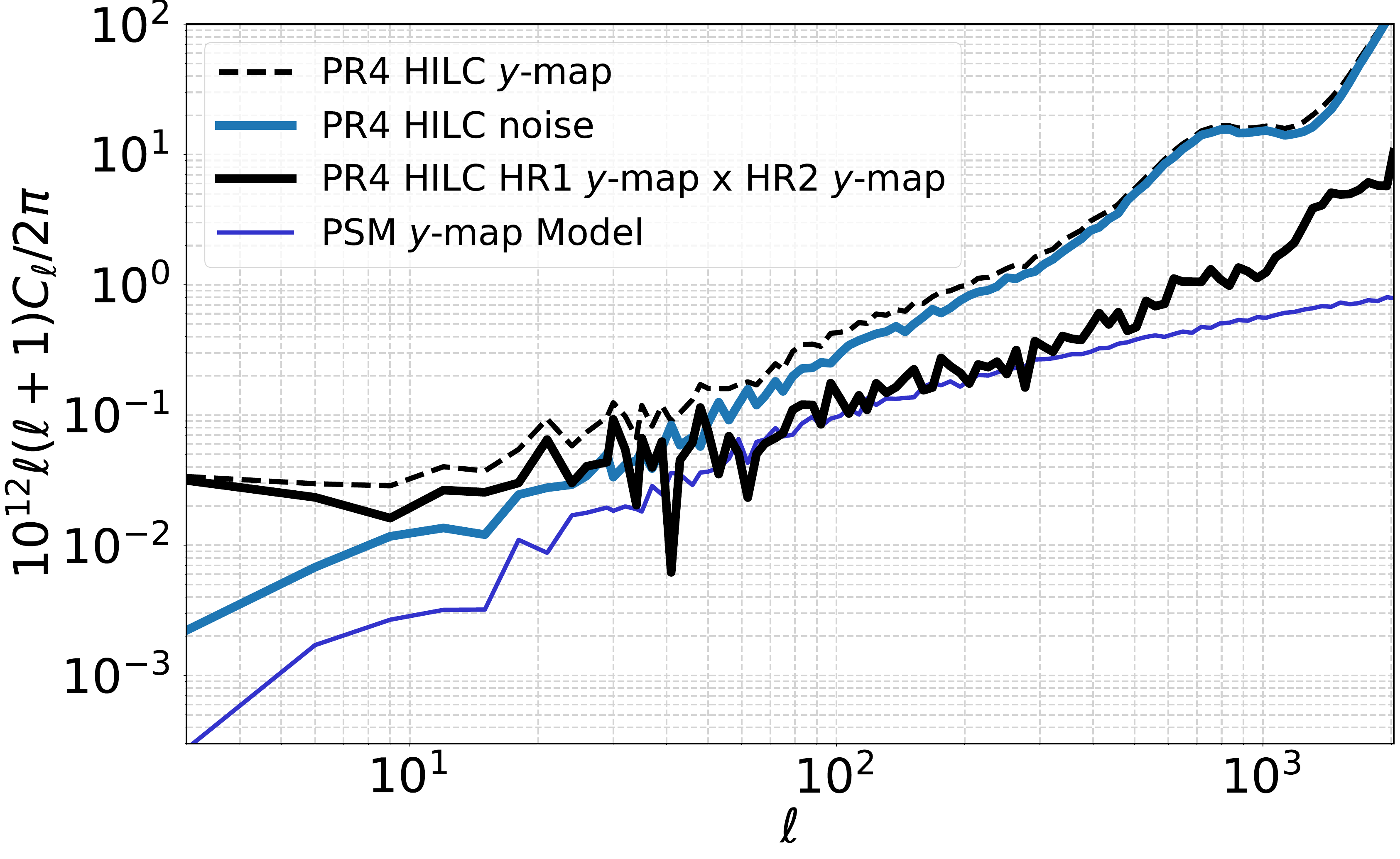}
    \caption{Results from ILC in harmonic domain (HILC): reconstructed thermal SZ $y$-map (top) and power spectra (bottom). The HILC $y$-map is much noisier compared to the NILC $y$-map and shows more significant residual excess power at both low and high multipoles.}
    \label{fig:tSZ_hilc}
\end{figure}

Fig.~\ref{fig:tSZ_hilc} (top panel) shows the thermal SZ Compton $y$-map reconstructed with the HILC method using the \textit{Planck} PR4 data (HILC $y$-map). Except for needlet decomposition, all parameters used are similar to those of the NILC pipeline as described in  Section~\ref{sec:nilc_method}.
The maximum multipole till which a frequency channel is used in the HILC is determined by a cutoff multipole at which the instrumental beam window function of that channel drops to $10^{-3}$. 
As can be seen from the comparison between Fig.~\ref{fig:tSZ_hilc} and Fig.~\ref{fig:tSZ_orth_map_plk_npipe_pr2}, the HILC $y$-map is noticeably noisier than the NILC $y$-map.  This is confirmed by the 1-PDF of the $y$-maps in Fig.~\ref{fig:tSZ_1PDF} and the 1-PDF of the noise in Fig.~\ref{fig:tSZ_noise_1PDF}, both of which show significantly larger variance for the HILC than for the NILC. The unnormalized skewness of the HILC $y$-map is also larger than that of the NILC $y$-map even after masking with GAL-MASK and PS-MASK. This points towards greater IR source contamination in the former map due to insufficient spatial localization from the HILC. All of this results in the NILC $y$-map having a larger signal-to-noise than the HILC $y$-map for thermal SZ observations.

The power spectrum of the HILC $y$-map is shown in the bottom panel of Fig.~\ref{fig:tSZ_hilc}, along with the power spectrum of the noise associated with the HILC $y$-map as obtained from PR4 half-ring data splits. The HILC $y$-map power spectrum obtained from the cross-correlation of half-ring maps is also shown (thick black line). In this plot, the excess due to the noise is clearly visible at high multipoles and low multipoles in comparison with the PR4 NILC $y$-map power spectrum. When considering the case of the cross-spectrum of half-ring maps, which is free from noise bias, the HILC $y$-map power spectrum still shows some excess power at high multipoles compared to NILC, which indicates larger contamination from extragalactic foregrounds. Similarly, at very large scales, an additional excess not accounted for by the noise is also visible, which indicates stronger large-scale residual contamination from Galactic foregrounds in the HILC $y$-map than in the NILC $y$-map. This again is due to the fact that HILC lacks the spatial localization to deal efficiently with inhomogeneous and anisotropic foregrounds. 

The average signal-to-noise ratio (SNR) is computed over the multipole range $\ell = 30$-$1000$ for both HILC and NILC $y$-maps as
\begin{equation} 
    \left(\frac{S}{N}\right) = \sqrt{\sum_{\ell=\ell_{\rm min}}^{\ell_{\rm max}} \left(\frac{C_\ell^{\text{HR1}\times \text{HR2}}}{\sigma_\ell}\right)^2}\,,
\end{equation}
where $C_\ell^{\text{HR1}\times \text{HR2}}$ is the cross-spectrum of HR1 and HR2 $y$-maps and  the uncertainty is derived from the auto-spectrum $C_\ell$ as
\begin{equation} 
    \sigma_\ell = \sqrt{2\over (2\ell+1)f_{\rm sky}\Delta\ell}\,C_\ell
\end{equation}
to account for the sample variance of the signal, residual foregrounds and noise. The SNR is $61.2$ for HILC and $178.2$ for NILC. This indicates a factor of $3$ improvement in SNR for the NILC $y$-map compared to the HILC $y$-map.

%%%%%%%%%%%%%%%%%%%%%%%%%%%%%%%%%%%%%%%%%%%%%%%%%%

% Don't change these lines
\bsp	% typesetting comment
\label{lastpage}

\end{document}

%% file: Planck.tex
\def\setsymbol#1#2{\expandafter\def\csname #1\endcsname{#2}}
\def\getsymbol#1{\csname #1\endcsname}

\newbox\tablebox    \newdimen\tablewidth
\def\leaderfil{\leaders\hbox to 5pt{\hss.\hss}\hfil}
\def\tablenote#1 #2\par{\begingroup \parindent=0.8em
    \abovedisplayshortskip=0pt\belowdisplayshortskip=0pt
    \noindent
    $$\hss\vbox{\hsize\tablewidth \hangindent=\parindent \hangafter=1 \noindent
    \hbox to \parindent{$^#1$\hss}\strut#2\strut\par}\hss$$
    \endgroup}

\def\L2{\ifmmode L_2\else $L_2$\fi}

\def\DeltaT{\ifmmode \Delta T\else $\Delta T$\fi}
\def\deltat{\ifmmode \Delta t\else $\Delta t$\fi}
\def\fknee{\ifmmode f_{\rm knee}\else $f_{\rm knee}$\fi}
\def\Fmax{\ifmmode F_{\rm max}\else $F_{\rm max}$\fi}
\def\solar{\ifmmode{\rm M}_{\mathord\odot}\else${\rm M}_{\mathord\odot}$\fi}
\def\Msolar{\ifmmode{\rm M}_{\mathord\odot}\else${\rm M}_{\mathord\odot}$\fi}
\def\Lsolar{\ifmmode{\rm L}_{\mathord\odot}\else${\rm L}_{\mathord\odot}$\fi}
\def\inv{\ifmmode^{-1}\else$^{-1}$\fi}
\def\mo{\ifmmode^{-1}\else$^{-1}$\fi}
\def\sup#1{\ifmmode ^{\rm #1}\else $^{\rm #1}$\fi}
\def\expo#1{\ifmmode \times 10^{#1}\else $\times 10^{#1}$\fi}
\def\,{\thinspace}
\def\lsim{\mathrel{\raise .4ex\hbox{\rlap{$<$}\lower 1.2ex\hbox{$\sim$}}}}
\def\gsim{\mathrel{\raise .4ex\hbox{\rlap{$>$}\lower 1.2ex\hbox{$\sim$}}}}

\def\simprop{\mathrel{\raise .4ex\hbox{\rlap{$\propto$}\lower 1.2ex\hbox{$\sim$}}}}
\def\deg{\ifmmode^\circ\else$^\circ$\fi}
\def\pdeg{\ifmmode $\setbox0=\hbox{$^{\circ}$}\rlap{\hskip.11\wd0 .}$^{\circ}
          \else \setbox0=\hbox{$^{\circ}$}\rlap{\hskip.11\wd0 .}$^{\circ}$\fi}
\def\arcs{\ifmmode {^{\scriptstyle\prime\prime}}
          \else $^{\scriptstyle\prime\prime}$\fi}
\def\arcm{\ifmmode {^{\scriptstyle\prime}}
          \else $^{\scriptstyle\prime}$\fi}
\newdimen\sa  \newdimen\sb
\def\parcs{\sa=.07em \sb=.03em
     \ifmmode \hbox{\rlap{.}}^{\scriptstyle\prime\kern -\sb\prime}\hbox{\kern -\sa}
     \else \rlap{.}$^{\scriptstyle\prime\kern -\sb\prime}$\kern -\sa\fi}
\def\parcm{\sa=.08em \sb=.03em
     \ifmmode \hbox{\rlap{.}\kern\sa}^{\scriptstyle\prime}\hbox{\kern-\sb}
     \else \rlap{.}\kern\sa$^{\scriptstyle\prime}$\kern-\sb\fi}
\def\ra[#1 #2 #3.#4]{#1\sup{h}#2\sup{m}#3\sup{s}\llap.#4}
\def\dec[#1 #2 #3.#4]{#1\deg#2\arcm#3\arcs\llap.#4}
\def\deco[#1 #2 #3]{#1\deg#2\arcm#3\arcs}
\def\rra[#1 #2]{#1\sup{h}#2\sup{m}}

\def\dots{\relax\ifmmode \ldots\else $\ldots$\fi}
\def\WHzsr{\ifmmode $W\,Hz\mo\,sr\mo$\else W\,Hz\mo\,sr\mo\fi}
\def\mHz{\ifmmode $\,mHz$\else \,mHz\fi}
\def\GHz{\ifmmode $\,GHz$\else \,GHz\fi}
\def\mKs{\ifmmode $\,mK\,s$^{1/2}\else \,mK\,s$^{1/2}$\fi}
\def\muKs{\ifmmode \,\mu$K\,s$^{1/2}\else \,$\mu$K\,s$^{1/2}$\fi}
\def\muKRJs{\ifmmode \,\mu$K$_{\rm RJ}$\,s$^{1/2}\else \,$\mu$K$_{\rm RJ}$\,s$^{1/2}$\fi}
\def\muKHz{\ifmmode \,\mu$K\,Hz$^{-1/2}\else \,$\mu$K\,Hz$^{-1/2}$\fi}
\def\MJysr{\ifmmode \,$MJy\,sr\mo$\else \,MJy\,sr\mo\fi}
\def\MJysrmK{\ifmmode \,$MJy\,sr\mo$\,mK$_{\rm CMB}\mo\else \,MJy\,sr\mo\,mK$_{\rm CMB}\mo$\fi}
\def\microns{\ifmmode \,\mu$m$\else \,$\mu$m\fi}
\def\micron{\microns}
\def\muK{\ifmmode \,\mu$K$\else \,$\mu$\hbox{K}\fi}
\def\microK{\ifmmode \,\mu$K$\else \,$\mu$\hbox{K}\fi}
\def\muW{\ifmmode \,\mu$W$\else \,$\mu$\hbox{W}\fi}
\def\kms{\ifmmode $\,km\,s$^{-1}\else \,km\,s$^{-1}$\fi}
\def\kmsMpc{\ifmmode $\,\kms\,Mpc\mo$\else \,\kms\,Mpc\mo\fi}
\providecommand{\sorthelp}[1]{}